\documentclass[a4paper,11pt]{article}
\usepackage{jheppub} 
\usepackage[T1]{fontenc} 
\usepackage{amsmath,amsthm}
\usepackage{amssymb, amsfonts}
\usepackage{bbm}

\usepackage{dcolumn}
\usepackage{bm}

\usepackage[T1]{fontenc} 
\usepackage{amsmath,amsthm}
\usepackage{amssymb, amsfonts}
\usepackage{bbm}
\usepackage{graphicx}
\usepackage{dsfont}
\usepackage{relsize}
\usepackage{stmaryrd}
\usepackage{lmodern}
\usepackage{slantsc}
\usepackage{scalefnt}
\usepackage{wasysym}
\usepackage{xspace}
\usepackage{boxedminipage}
\usepackage{ifpdf}
\usepackage{multirow, hhline}
\usepackage{slashed}
\usepackage{xifthen}
\usepackage{tensor}
\usepackage{array, nicematrix}
\usepackage{tabu}
\usepackage{pbox}
\usepackage{multirow}
\usepackage{tikz}
\usepackage{overpic}
\usepackage{color}
\usepackage{diagbox}
\usepackage{makecell}
\usetikzlibrary{arrows, matrix, calc, scopes, decorations.markings}
\usepackage[mathcal]{euscript}
\usepackage{booktabs}
\usepackage{autobreak, subfig, diagbox, mathrsfs}
\usepackage{mathbbol}
\usepackage{tcolorbox}
\usepackage{hyperref}
\usepackage{adjustbox}
\allowdisplaybreaks[4]

\newcommand{\NN}{\mathbb{N}}
\newcommand{\Z}{\mathbb{Z}}

\newcommand{\M}{\mathcal{M}}
\newcommand{\Hil}{\mathcal{H}}

\newcommand{\Cent}{\mathcal{Z}}
\newcommand{\T}{\mathcal{T}}

\newcommand{\C}{\mathbb{C}}

\newcommand{\Fus}{\mathscr{F}}

\newcommand{\RR}{\mathbb{R}}
\newcommand{\CC}{\mathbb{C}}
\newcommand{\idm}{{\mathds{1}}}
\newcommand{\nulm}{{\mathbb{O}}}

\newcommand{\df}{{\mathrm{d}}}

\newcommand{\dk}[2][1]{{\ifthenelse{\equal{#1}{1}}{\frac{\df{#2}}{2\pi}}{\frac{\df^{#1}{#2}}{(2\pi)^{#1}}}}}

\renewcommand{\Vec}{{\tt Vec}}

\newcommand{\eqs}[1]{{\texorpdfstring{${#1}$}{Lg}}}

\newcommand{\ket}[1]{{\left\vert{#1}\right\rangle}}

\newcommand{\eq}[1]{\begin{align*}#1\end{align*}}
\newcommand{\eqn}[2][0]{\ifthenelse{\equal{#1}{0}}{\begin{equation}\begin{aligned}#2\end{aligned}\end{equation}}{\begin{equation}\begin{aligned}#2\end{aligned}\label{#1}\end{equation}}}

\usetikzlibrary{shapes.geometric, snakes}
\tikzset{>=latex}
\usetikzlibrary{arrows, matrix, calc, scopes, decorations.markings}
\usetikzlibrary{decorations.pathmorphing, positioning}
\tikzset{snake it/.style={decorate, decoration={snake,amplitude=0.2mm,segment length=1mm}}}
\tikzset{->-/.style={decoration={
			 markings,
			 mark=at position .5*\pgfdecoratedpathlength+2pt with {\arrow{>}}},postaction={decorate}}}

\tikzset{-<-/.style={decoration={
			 markings,
			 mark=at position .5*\pgfdecoratedpathlength+2pt with {\arrow{<}}},postaction={decorate}}}

\newtabulinestyle{dash=off 2pt}

\renewcommand{\arraystretch}{1.5}

{\null\,\vcenter\bgroup\scriptsize
\renewcommand{\arraystretch}{0.7}%
\arraycolsep=.13885em
\hbox\bgroup$\array{@{}#1@{}}}
{\endarray$\egroup\egroup\,\null}
\newcommand{\Lattice}{
\begin{tikzpicture}[baseline={([yshift=-.8ex]current bounding box.center)},scale=1.]
\draw [decorate,decoration={snake,segment length=.12cm, amplitude=.04cm}] (1.7,6.8) -- (2.3,6.8);
\draw [] (1.7,6.4) -- (1.7,6.8);
\draw [] (1.7,6.8) -- (1.7,7.2);
\draw [] (1.7,7.2) -- (2.4,7.6);
\draw [] (2.4,7.6) -- (3.1,7.2);
\draw [] (2.4,6.) -- (1.7,6.4);
\draw [] (3.1,6.4) -- (2.4,6.);
\draw [] (2.4,5.2) -- (2.4,5.6);
\draw [] (2.4,5.6) -- (2.4,6.);
\draw [decorate,decoration={snake,segment length=.12cm, amplitude=.04cm}] (2.4,5.6) -- (3.,5.6);
\draw [decorate,decoration={snake,segment length=.12cm, amplitude=.04cm}] (3.1,6.8) -- (3.7,6.8);
\draw [] (3.1,6.4) -- (3.1,6.8);
\draw [] (3.1,6.8) -- (3.1,7.2);
\draw [] (3.1,7.2) -- (3.8,7.6);
\draw [] (3.8,7.6) -- (4.5,7.2);
\draw [] (3.8,6.) -- (3.1,6.4);
\draw [] (4.5,6.4) -- (3.8,6.);
\draw [] (3.8,5.2) -- (3.8,5.6);
\draw [] (3.8,5.6) -- (3.8,6.);
\draw [decorate,decoration={snake,segment length=.12cm, amplitude=.04cm}] (3.8,5.6) -- (4.4,5.6);
\draw [decorate,decoration={snake,segment length=.12cm, amplitude=.04cm}] (4.5,6.8) -- (5.1,6.8);
\draw [] (4.5,6.4) -- (4.5,6.8);
\draw [] (4.5,6.8) -- (4.5,7.2);
\draw [] (4.5,7.2) -- (5.2,7.6);
\draw [] (5.2,7.6) -- (5.9,7.2);
\draw [] (5.2,6.) -- (4.5,6.4);
\draw [] (5.9,6.4) -- (5.2,6.);
\draw [] (5.2,5.2) -- (5.2,5.6);
\draw [] (5.2,5.6) -- (5.2,6.);
\draw [decorate,decoration={snake,segment length=.12cm, amplitude=.04cm}] (5.2,5.6) -- (5.8,5.6);
\draw [decorate,decoration={snake,segment length=.12cm, amplitude=.04cm}] (5.9,6.8) -- (6.5,6.8);
\draw [] (5.9,6.4) -- (5.9,6.8);
\draw [] (5.9,6.8) -- (5.9,7.2);
\draw [] (5.9,7.2) -- (6.6,7.6);
\draw [] (6.6,7.6) -- (7.3,7.2);
\draw [] (6.6,6.) -- (5.9,6.4);
\draw [] (7.3,6.4) -- (6.6,6.);
\draw [] (6.6,5.2) -- (6.6,5.6);
\draw [] (6.6,5.6) -- (6.6,6.);
\draw [decorate,decoration={snake,segment length=.12cm, amplitude=.04cm}] (6.6,5.6) -- (7.2,5.6);
\draw [decorate,decoration={snake,segment length=.12cm, amplitude=.04cm}] (1.7,4.4) -- (2.3,4.4);
\draw [] (1.7,4.) -- (1.7,4.4);
\draw [] (1.7,4.4) -- (1.7,4.8);
\draw [] (1.7,4.8) -- (2.4,5.2);
\draw [] (2.4,5.2) -- (3.1,4.8);
\draw [] (2.4,3.6) -- (1.7,4.);
\draw [] (3.1,4.) -- (2.4,3.6);
\draw [] (2.4,3.2) -- (2.4,3.6);
\draw [decorate,decoration={snake,segment length=.12cm, amplitude=.04cm}] (3.1,4.4) -- (3.7,4.4);
\draw [] (3.1,4.) -- (3.1,4.4);
\draw [] (3.1,4.4) -- (3.1,4.8);
\draw [] (3.1,4.8) -- (3.8,5.2);
\draw [] (3.8,5.2) -- (4.5,4.8);
\draw [] (3.8,3.6) -- (3.1,4.);
\draw [] (4.5,4.) -- (3.8,3.6);
\draw [] (3.8,3.2) -- (3.8,3.6);
\draw [decorate,decoration={snake,segment length=.12cm, amplitude=.04cm}] (4.5,4.4) -- (5.1,4.4);
\draw [] (4.5,4.) -- (4.5,4.4);
\draw [] (4.5,4.4) -- (4.5,4.8);
\draw [] (4.5,4.8) -- (5.2,5.2);
\draw [] (5.2,5.2) -- (5.9,4.8);
\draw [] (5.2,3.6) -- (4.5,4.);
\draw [] (5.9,4.) -- (5.2,3.6);
\draw [] (5.2,3.2) -- (5.2,3.6);
\draw [decorate,decoration={snake,segment length=.12cm, amplitude=.04cm}] (5.9,4.4) -- (6.5,4.4);
\draw [] (5.9,4.) -- (5.9,4.4);
\draw [] (5.9,4.4) -- (5.9,4.8);
\draw [] (5.9,4.8) -- (6.6,5.2);
\draw [] (6.6,5.2) -- (7.3,4.8);
\draw [] (6.6,3.6) -- (5.9,4.);
\draw [] (7.3,4.) -- (6.6,3.6);
\draw [] (6.6,3.2) -- (6.6,3.6);
\draw [decorate,decoration={snake,segment length=.12cm, amplitude=.04cm}] (7.3,6.8) -- (7.9,6.8);
\draw [] (7.3,6.4) -- (7.3,6.8);
\draw [] (7.3,6.8) -- (7.3,7.2);
\draw [] (7.3,7.2) -- (7.6,7.4);
\draw [] (8.,6.) -- (7.3,6.4);
\draw [] (8.3,6.2) -- (8.,6.);
\draw [] (8.,5.2) -- (8.,5.6);
\draw [] (8.,5.6) -- (8.,6.);
\draw [decorate,decoration={snake,segment length=.12cm, amplitude=.04cm}] (8.,5.6) -- (8.6,5.6);
\draw [decorate,decoration={snake,segment length=.12cm, amplitude=.04cm}] (7.3,4.4) -- (7.9,4.4);
\draw [] (7.3,4.) -- (7.3,4.4);
\draw [] (7.3,4.4) -- (7.3,4.8);
\draw [] (7.3,4.8) -- (8.,5.2);
\draw [] (8.,5.2) -- (8.3,5.);
\draw [] (8.,3.6) -- (7.3,4.);
\draw [] (8.,3.2) -- (8.,3.6);
\draw [] (6.6,7.6) -- (6.6,8.);
\draw [] (5.2,7.6) -- (5.2,8.);
\draw [] (3.8,7.6) -- (3.8,8.);
\draw [] (2.4,7.6) -- (2.4,8.);
\draw [] (8.3,3.8) -- (8.,3.6);
\draw [] (1.7,4.) -- (1.3,3.8);
\draw [] (1.7,4.8) -- (1.3,5.);
\draw [] (1.7,6.4) -- (1.3,6.2);
\draw [] (1.7,7.2) -- (1.3,7.4);
\end{tikzpicture}
}

\newcommand{\FatLattice}{
\begin{tikzpicture}[baseline={([yshift=-.8ex]current bounding box.center)},scale=.85]
\draw [] (0.1,0.057) -- (0.1,0.9);
\draw [] (-0.1,0.057) -- (-0.1,1.943);
\draw [] (0.1,1.1) -- (0.1,1.943);
\draw [] (0.1,1.943) -- (1.714,2.886);
\draw [] (0.1,0.057) -- (1.714,-0.886);
\draw [] (1.714,2.886) -- (3.328,1.943);
\draw [] (1.714,-0.886) -- (3.328,0.057);
\draw [] (3.328,1.943) -- (3.328,0.057);
\draw [] (-0.1,0.057) -- (-0.907,-0.414);
\draw [] (-0.1,1.943) -- (-0.907,2.414);
\draw [] (0.,-0.114) -- (-0.807,-0.586);
\draw [] (0.,2.114) -- (-0.807,2.586);
\draw [] (0.,2.114) -- (1.614,3.057);
\draw [] (1.614,3.057) -- (1.614,4.057);
\draw [] (1.814,3.057) -- (1.814,4.057);
\draw [] (1.814,3.057) -- (3.428,2.114);
\draw [] (3.428,2.114) -- (4.235,2.586);
\draw [] (3.528,1.943) -- (4.335,2.414);
\draw [] (0.,-0.114) -- (1.614,-1.057);
\draw [] (1.614,-1.057) -- (1.614,-2.057);
\draw [] (1.814,-1.057) -- (1.814,-2.057);
\draw [] (1.814,-1.057) -- (3.428,-0.114);
\draw [] (3.428,-0.114) -- (4.235,-0.586);
\draw [] (3.528,0.057) -- (4.335,-0.414);
\draw [] (3.528,0.057) -- (3.528,0.9);
\draw [] (3.528,1.943) -- (3.528,1.1);
\draw [decorate,decoration={snake,segment length=.12cm, amplitude=.04cm}] (0.1,0.9) -- (1.7,0.9);
\draw [decorate,decoration={snake,segment length=.12cm, amplitude=.04cm}] (0.1,1.1) -- (1.7,1.1);
\draw [decorate,decoration={snake,segment length=.12cm, amplitude=.04cm}] (3.528,0.9) -- (4.2,0.9);
\draw [decorate,decoration={snake,segment length=.12cm, amplitude=.04cm}] (3.528,1.1) -- (4.2,1.1);

\node [centered, rotate=0.] at (0.,0.5) {\scriptsize $x$};
\node [centered, rotate=0.] at (0.,1.5) {\scriptsize $y$};
\node [centered, rotate=0.] at (-0.3,0.5) {\scriptsize $g_i$};
\node [centered, rotate=0.] at (0.3,0.5) {\scriptsize $g_j$};
\node [centered, rotate=0.] at (-0.3,1.5) {\scriptsize $g_i$};
\node [centered, rotate=0.] at (0.3,1.5) {\scriptsize $g_j$};
\node [centered, rotate=0.] at (1.,1.) {\scriptsize $z$};
\node [centered, rotate=0.] at (1.,1.3) {\scriptsize $g_j$};
\node [centered, rotate=0.] at (1.,0.7) {\scriptsize $g_j$};
\node [centered, rotate=0.] at (0.857,2.5) {\scriptsize $u$};
\node [centered, rotate=0.] at (0.715,2.7) {\scriptsize $g_k$};
\node [centered, rotate=0.] at (1.045,2.3) {\scriptsize $g_j$};
\node [centered, rotate=0.] at (2.571,2.5) {\scriptsize $v$};
\node [centered, rotate=0.] at (2.421,2.3) {\scriptsize $g_j$};
\node [centered, rotate=0.] at (2.631,2.75) {\scriptsize $g_l$};
\end{tikzpicture}
}

\newcommand{\SymmTrans}{
\begin{tikzpicture}[baseline={([yshift=-.8ex]current bounding box.center)},scale=.6]
\draw [->, color=red] (-1.,0.) .. controls (-1.,-2.) and (1.,-2.) .. (1.,-4.) ;
\draw [color=blue, ultra thick] (-2.,-2.)--(2.,-2.);
\draw [color=blue, ultra thick] (-2.,-2.)--(-3,-0.3);
\draw [color=blue, ultra thick] (2.,-2.)--(3.,-3.7);
\draw [] (-2.,-2.)--(-3.,-3.7);
\draw [] (2.,-2.)--(3,-0.3);

\node [centered, rotate=0.] at (-.5,0.) {$a$};
\node [centered, rotate=0.] at (.0,-3.75) {$\mathcal{G}(a)$};
\node [centered, rotate=0.] at (-3.8,0.) {DW};

\fill [red] (-1.,0.) circle (4pt);
\fill [red] (1.,-4.) circle (4pt);
\end{tikzpicture}
}

\newcommand{\DW}{
\begin{tikzpicture}[baseline={([yshift=-.8ex]current bounding box.center)},scale=.35]
\draw [->-, blue, ultra thick] (0.,0.)--(0.,4.);

\node [centered, rotate=0.] at (0.,-0.3) {\scriptsize $\alpha_{+-}$};
\node [centered, rotate=0.] at (-2.,2.) {$+$};
\node [centered, rotate=0.] at (2.,2.) {$-$};
\end{tikzpicture}
\qquad=\qquad
\begin{tikzpicture}[baseline={([yshift=-.8ex]current bounding box.center)},scale=.35]
\draw [->-, blue, ultra thick] (0.,4.)--(0.,0.);

\node [centered, rotate=0.] at (0.,-0.3) {\scriptsize $\alpha_{-+}$};
\node [centered, rotate=0.] at (-2.,2.) {$+$};
\node [centered, rotate=0.] at (2.,2.) {$-$};
\end{tikzpicture}
}

\newcommand{\PlaquetteSrc}{
\begin{tikzpicture}[baseline={([yshift=-.8ex]current bounding box.center)},scale=.5]
\draw [->-] (3.2,5.6) -- (3.2,6.8);
\draw [->-] (3.2,7.2) -- (4.6,8.);
\draw [->-] (4.6,8.) -- (6.,7.2);
\draw [->-] (6.,5.6) -- (4.6,4.8);
\draw [->-] (4.6,4.8) -- (3.2,5.6);
\draw [->-] (6.,7.2) -- (6.,5.6);
\draw [decorate,decoration={snake,segment length=.12cm, amplitude=.04cm}, ->] (3.2,6.4) -- (4.4,6.4);
\draw [->-] (2.6,7.6) -- (3.2,7.2);
\draw [->-] (4.6,8.8) -- (4.6,8.);
\draw [->-] (6.6,7.6) -- (6.,7.2);
\draw [->-] (6.6,5.2) -- (6.,5.6);
\draw [->-] (4.6,4.) -- (4.6,4.8);
\draw [->-] (2.6,5.2) -- (3.2,5.6);
\draw [->-] (3.2,6.6) -- (3.2,7.2);
\node [centered, rotate=0.] at (2.4,7.8) {\scriptsize $e_1$};
\node [centered, rotate=0.] at (4.95,8.8) {\scriptsize $e_2$};
\node [centered, rotate=0.] at (6.9,7.8) {\scriptsize $e_3$};
\node [centered, rotate=0.] at (6.9,5.) {\scriptsize $e_4$};
\node [centered, rotate=0.] at (5.,4.) {\scriptsize $e_5$};
\node [centered, rotate=0.] at (2.3,5.) {\scriptsize $e_6$};
\node [centered, rotate=0.] at (2.85,6.) {\scriptsize $i_7$};
\node [centered, rotate=0.] at (2.85,6.9) {\scriptsize $i_1$};
\node [centered, rotate=0.] at (3.6,7.9) {\scriptsize $i_2$};
\node [centered, rotate=0.] at (5.6,7.9) {\scriptsize $i_3$};
\node [centered, rotate=0.] at (6.4,6.4) {\scriptsize $i_4$};
\node [centered, rotate=0.] at (5.6,4.9) {\scriptsize $i_5$};
\node [centered, rotate=0.] at (3.7,4.9) {\scriptsize $i_6$};
\node [centered, rotate=0.] at (4.7,6.4) {\scriptsize $p$};
\end{tikzpicture}
}

\newcommand{\PlaquetteTar}{
\begin{tikzpicture}[baseline={([yshift=-.8ex]current bounding box.center)},scale=.5]
\draw [->-] (3.2,5.6) -- (3.2,6.8);
\draw [->-] (3.2,7.2) -- (4.6,8.);
\draw [->-] (4.6,8.) -- (6.,7.2);
\draw [->-] (6.,5.6) -- (4.6,4.8);
\draw [->-] (4.6,4.8) -- (3.2,5.6);
\draw [->-] (6.,7.2) -- (6.,5.6);
\draw [decorate,decoration={snake,segment length=.12cm, amplitude=.04cm}] (3.2,6.4) -- (4.4,6.4);
\draw [->-] (2.6,7.6) -- (3.2,7.2);
\draw [->-] (4.6,8.8) -- (4.6,8.);
\draw [->-] (6.6,7.6) -- (6.,7.2);
\draw [->-] (6.6,5.2) -- (6.,5.6);
\draw [->-] (4.6,4.) -- (4.6,4.8);
\draw [->-] (2.6,5.2) -- (3.2,5.6);
\draw [->-] (3.2,6.6) -- (3.2,7.2);
\node [centered, rotate=0.] at (2.4,7.8) {\scriptsize $e_1$};
\node [centered, rotate=0.] at (4.95,8.8) {\scriptsize $e_2$};
\node [centered, rotate=0.] at (6.9,7.8) {\scriptsize $e_3$};
\node [centered, rotate=0.] at (6.9,5.) {\scriptsize $e_4$};
\node [centered, rotate=0.] at (5.,4.) {\scriptsize $e_5$};
\node [centered, rotate=0.] at (2.3,5.) {\scriptsize $e_6$};
\node [centered, rotate=0.] at (2.85,6.) {\scriptsize $j_7$};
\node [centered, rotate=0.] at (2.85,6.9) {\scriptsize $j_1$};
\node [centered, rotate=0.] at (3.6,7.9) {\scriptsize $j_2$};
\node [centered, rotate=0.] at (5.6,7.9) {\scriptsize $j_3$};
\node [centered, rotate=0.] at (6.4,6.4) {\scriptsize $j_4$};
\node [centered, rotate=0.] at (5.6,4.9) {\scriptsize $j_5$};
\node [centered, rotate=0.] at (3.7,4.9) {\scriptsize $j_6$};
\node [centered, rotate=0.] at (4.7,6.4) {\scriptsize $1$};
\end{tikzpicture}
}

\newcommand{\PachnerOne}{\begin{tikzpicture}[baseline={([yshift=-.8ex]current bounding box.center)},scale=.6]
\draw [->-] (0.,0.2) -- (.5,1.);
\draw [->-] (0.,1.8) -- (.5,1.);
\draw [->-] (2.,1.) -- (.5,1.);
\draw [->-] (2.5,1.8) -- (2.,1.);
\draw [->-] (2.5,0.2) -- (2.,1.);
\node at (1.25, 1.3) {\scriptsize $m$};
\node at (-.15, 2.) {\scriptsize $a$};
\node at (-.15, 0.) {\scriptsize $b$};
\node at (2.65, 2.) {\scriptsize $d$};
\node at (2.65, 0.) {\scriptsize $c$};
\end{tikzpicture}\qquad = \qquad \sum_{n\in L_{\Fus}}\ \sqrt{d_md_n}\ G^{abm}_{cdn}\quad \begin{tikzpicture}[baseline={([yshift=-.8ex]current bounding box.center)},scale=.6]
\draw [->-] (0.,0.) -- (.8,.5);
\draw [->-] (1.6,0.) -- (.8,.5);
\draw [->-] (.8,.5) -- (.8,2.);
\draw [->-] (0.,2.5) -- (.8,2.);
\draw [->-] (1.6,2.5) -- (.8,2.);
\node at (1.05, 1.3) {\scriptsize $n$};
\node at (-.15, 2.6) {\scriptsize $a$};
\node at (-.15, -.1) {\scriptsize $b$};
\node at (1.75, 2.6) {\scriptsize $d$};
\node at (1.75, -0.1) {\scriptsize $c$};
\end{tikzpicture}}

\newcommand{\PachnerTwo}{\begin{tikzpicture}[baseline={([yshift=-.8ex]current bounding box.center)},scale=.6]
\draw [->-] (1, 0) -- (1, 1);
\draw [->-] (1, 1) arc (270:90:.5);
\draw [->-] (1, 1) arc (-90:90:.5);
\draw [->-] (1, 2) -- (1, 3);
\node at (1.25, .3) {\scriptsize $a$};
\node at (1., 1.5) {\scriptsize \color{red}$\times$};
\node at (.2, 1.5) {\scriptsize $x$};
\node at (1.8, 1.5) {\scriptsize $y$};
\node at (1.25, 2.7) {\scriptsize $b$};
\end{tikzpicture}\qquad = \qquad \sqrt{\frac{d_xd_y}{d_a}}\ \delta_{ab}\ \delta_{xya^\ast}\quad \begin{tikzpicture}[baseline={([yshift=-.8ex]current bounding box.center)},scale=.5]
\draw [->-] (1, 0) -- (1, 3);
\node at (1.25, 1.5) {\scriptsize $a$};
\end{tikzpicture}}

\newcommand{\PachnerThree}{\begin{tikzpicture}[baseline={([yshift=-.8ex]current bounding box.center)},scale=.5]
\draw [->-] (1, 0) -- (1, 3);
\node at (1.25, 1.5) {\scriptsize $a$};
\end{tikzpicture}\quad = \quad \frac{1}{D}\sum_{xy\in L_{\Fus}} \sqrt{\frac{d_xd_y}{d_a}}\ \delta_{xya^\ast} \quad \begin{tikzpicture}[baseline={([yshift=-.8ex]current bounding box.center)},scale=.6]
\draw [->-] (1, 0) -- (1, 1);
\draw [->-] (1, 1) arc (270:90:.5);
\draw [->-] (1, 1) arc (-90:90:.5);
\draw [->-] (1, 2) -- (1, 3);
\node at (1.25, .3) {\scriptsize $a$};
\node at (.2, 1.5) {\scriptsize $x$};
\node at (1.8, 1.5) {\scriptsize $y$};
\node at (1.25, 2.7) {\scriptsize $a$};
\end{tikzpicture}}

\newcommand{\PachnerFour}{
\begin{tikzpicture}[baseline={([yshift=-.8ex]current bounding box.center)},scale=.6]
\draw [->-] (3.2,5.6) -- (3.2,6.8);
\draw [->-] (3.2,7.2) -- (4.6,8.);
\draw [->-] (4.6,8.) -- (6.,7.2);
\draw [->-] (6.,5.6) -- (4.6,4.8);
\draw [->-] (4.6,4.8) -- (3.2,5.6);
\draw [->-] (6.,7.2) -- (6.,5.6);
\draw [decorate,decoration={snake,segment length=.12cm, amplitude=.04cm}, ->] (3.2,6.6) -- (4.4,6.6);
\draw [->-] (2.6,7.6) -- (3.2,7.2);
\draw [->-] (4.6,8.8) -- (4.6,8.);
\draw [->-] (6.6,7.6) -- (6.,7.2);
\draw [->-] (6.6,5.2) -- (6.,5.6);
\draw [->-] (4.6,4.) -- (4.6,4.8);
\draw [->-] (2.6,5.2) -- (3.2,5.6);
\draw [->-] (3.2,6.6) -- (3.2,7.2);
\node [centered, rotate=0.] at (2.5,7.8) {\scriptsize $e_1$};
\node [centered, rotate=0.] at (2.9,6.) {\scriptsize $i_0$};
\node [centered, rotate=0.] at (2.9,6.9) {\scriptsize $i_1$};
\node [centered, rotate=0.] at (3.6,7.8) {\scriptsize $i_2$};
\node [centered, rotate=0.] at (4.7,6.6) {\scriptsize $p$};
\end{tikzpicture}\ =\quad\sum_{j\in L_\Fus}\sqrt{d_{i_1}d_j}\ G^{i_2^\ast e_1i_1}_{i_0 p^\ast j}\ 
\begin{tikzpicture}[baseline={([yshift=-.8ex]current bounding box.center)},scale=.6]
\draw [->-] (3.2,5.6) -- (3.2,7.2);
\draw [->-] (3.2,7.2) -- (3.55,7.4);
\draw [->-] (3.55,7.4) -- (4.6,8.);
\draw [->-] (4.6,8.) -- (6.,7.2);
\draw [->-] (6.,5.6) -- (4.6,4.8);
\draw [->-] (4.6,4.8) -- (3.2,5.6);
\draw [->-] (6.,7.2) -- (6.,5.6);
\draw [->-] (2.6,7.6) -- (3.2,7.2);
\draw [->-] (4.6,8.8) -- (4.6,8.);
\draw [->-] (6.6,7.6) -- (6.,7.2);
\draw [->-] (6.6,5.2) -- (6.,5.6);
\draw [->-] (4.6,4.) -- (4.6,4.8);
\draw [->-] (2.6,5.2) -- (3.2,5.6);
\draw [decorate,decoration={snake,segment length=.12cm, amplitude=.04cm}, ->] (3.59,7.42) -- (4.1,6.5);
\node [centered, rotate=0.] at (2.5,5.) {\scriptsize $e_n$};
\node [centered, rotate=0.] at (2.5,7.8) {\scriptsize $e_1$};
\node [centered, rotate=0.] at (2.9,6.5) {\scriptsize $i_0$};
\node [centered, rotate=0.] at (3.3,7.6) {\scriptsize $j$};
\node [centered, rotate=0.] at (3.9,8.) {\scriptsize $i_2$};
\node [centered, rotate=0.] at (4.3,6.3) {\scriptsize $p$};
\node [centered, rotate=0.] at (3.7,5.1) {\scriptsize $i_n$};
\end{tikzpicture}
}

\newcommand{\PachnerFive}{
\begin{tikzpicture}[baseline={([yshift=-.8ex]current bounding box.center)},scale=.6]
\draw [->-] (3.2,5.6) -- (3.2,6.1);
\draw [->-] (3.2,6.1) -- (3.2,6.7);
\draw [->-] (3.2,6.7) -- (3.2,7.2);
\draw [->-] (3.2,7.2) -- (4.6,8.);
\draw [->-] (4.6,8.) -- (6.,7.2);
\draw [->-] (6.,5.6) -- (4.6,4.8);
\draw [->-] (4.6,4.8) -- (3.2,5.6);
\draw [->-] (6.,7.2) -- (6.,5.6);
\draw [decorate,decoration={snake,segment length=.12cm, amplitude=.04cm}, ->] (3.2,6.1) -- (4.4,6.1);
\draw [decorate,decoration={snake,segment length=.12cm, amplitude=.04cm}, ->] (3.2,6.7) -- (4.4,6.7);
\draw [->-] (2.6,7.6) -- (3.2,7.2);
\draw [->-] (4.6,8.8) -- (4.6,8.);
\draw [->-] (6.6,7.6) -- (6.,7.2);
\draw [->-] (6.6,5.2) -- (6.,5.6);
\draw [->-] (4.6,4.) -- (4.6,4.8);
\draw [->-] (2.6,5.2) -- (3.2,5.6);
\node [centered, rotate=0.] at (2.9,5.8) {\scriptsize $i_0$};
\node [centered, rotate=0.] at (2.9,7.1) {\scriptsize $i_1$};
\node [centered, rotate=0.] at (4.7,6.7) {\scriptsize $r$};
\node [centered, rotate=0.] at (4.7,6.1) {\scriptsize $s$};
\node [centered, rotate=0.] at (2.9,6.4) {\scriptsize $j$};
\end{tikzpicture}\quad = \quad \sum_{u\in L_\Fus}\sqrt{d_jd_p}\ G^{r^\ast i_1^\ast j}_{i_0s^\ast p}\quad 
\begin{tikzpicture}[baseline={([yshift=-.8ex]current bounding box.center)},scale=.6]
\draw [->-] (3.2,5.6) -- (3.2,6.8);
\draw [->-] (3.2,7.2) -- (4.6,8.);
\draw [->-] (4.6,8.) -- (6.,7.2);
\draw [->-] (6.,5.6) -- (4.6,4.8);
\draw [->-] (4.6,4.8) -- (3.2,5.6);
\draw [->-] (6.,7.2) -- (6.,5.6);
\draw [decorate,decoration={snake,segment length=.12cm, amplitude=.04cm}, ->] (3.2,6.6) -- (4.4,6.6);
\draw [->-] (2.6,7.6) -- (3.2,7.2);
\draw [->-] (4.6,8.8) -- (4.6,8.);
\draw [->-] (6.6,7.6) -- (6.,7.2);
\draw [->-] (6.6,5.2) -- (6.,5.6);
\draw [->-] (4.6,4.) -- (4.6,4.8);
\draw [->-] (2.6,5.2) -- (3.2,5.6);
\draw [->-] (3.2,6.6) -- (3.2,7.2);
\node [centered, rotate=0.] at (2.9,6.) {\scriptsize $i_0$};
\node [centered, rotate=0.] at (2.9,6.9) {\scriptsize $i_1$};
\node [centered, rotate=0.] at (4.7,6.6) {\scriptsize $p$};
\end{tikzpicture}
}

\newcommand{\PlaquetteMsr}{
\begin{tikzpicture}[baseline={([yshift=-.8ex]current bounding box.center)},scale=.55]
\draw [->-] (3.2,5.6) -- (3.2,6.8);
\draw [->-] (3.2,7.2) -- (4.6,8.);
\draw [->-] (4.6,8.) -- (6.,7.2);
\draw [->-] (6.,5.6) -- (4.6,4.8);
\draw [->-] (4.6,4.8) -- (3.2,5.6);
\draw [->-] (6.,7.2) -- (6.,5.6);
\draw [->-] (2.6,7.6) -- (3.2,7.2);
\draw [->-] (4.6,8.8) -- (4.6,8.);
\draw [->-] (6.6,7.6) -- (6.,7.2);
\draw [->-] (6.6,5.2) -- (6.,5.6);
\draw [->-] (4.6,4.) -- (4.6,4.8);
\draw [->-] (2.6,5.2) -- (3.2,5.6);
\draw [->-] (3.2,6.6) -- (3.2,7.2);
\node [centered, rotate=0.] at (2.4,7.8) {\scriptsize $e_1$};
\node [centered, rotate=0.] at (4.95,8.8) {\scriptsize $e_2$};
\node [centered, rotate=0.] at (6.9,7.8) {\scriptsize $e_3$};
\node [centered, rotate=0.] at (6.9,5.) {\scriptsize $e_4$};
\node [centered, rotate=0.] at (5.,4.) {\scriptsize $e_5$};
\node [centered, rotate=0.] at (2.3,5.) {\scriptsize $e_6$};
\node [centered, rotate=0.] at (2.85,6.) {\scriptsize $i_7$};
\node [centered, rotate=0.] at (2.85,6.9) {\scriptsize $i_1$};
\node [centered, rotate=0.] at (3.6,7.9) {\scriptsize $i_2$};
\node [centered, rotate=0.] at (5.6,7.9) {\scriptsize $i_3$};
\node [centered, rotate=0.] at (6.4,6.4) {\scriptsize $i_4$};
\node [centered, rotate=0.] at (5.6,4.9) {\scriptsize $i_5$};
\node [centered, rotate=0.] at (3.7,4.9) {\scriptsize $i_6$};

\draw [decorate,decoration={snake,segment length=.12cm, amplitude=.04cm}, ->] (3.2,6.4) -- (5.,6.4);
\draw [->] (3.8, 6.4) arc (180:0:.8);
\draw [] (4.4, 6.4) arc (-180:0:0.5);
\node [centered, rotate=0.] at (4.8,6.7) {\scriptsize $p$};
\node [centered, rotate=0.] at (4.1,6.1) {\scriptsize $t$};
\node [centered, rotate=0.] at (3.5,6.1) {\scriptsize $p$};
\node [centered, rotate=0.] at (5.65,6.4) {\scriptsize $s$};
\node [centered, rotate=0.] at (4.6,5.4) {\color{red} $\times$};
\end{tikzpicture}
}

\newcommand{\Tail}[1]{
\begin{tikzpicture}[baseline={([yshift=-.8ex]current bounding box.center)},scale=1.]
\draw [decorate,decoration={snake,segment length=.12cm, amplitude=.04cm}] (0, 0) -- (0, 1.5);
\node [right, rotate=0.] at (0.,.75) {\scriptsize ${#1}$};
\end{tikzpicture}
}

\newcommand{\Composition}{   
\begin{tikzpicture}[baseline={([yshift=-.8ex]current bounding box.center)},scale=.5]

\draw[] (0.,0.) -- (0.,2.);
\draw[blue, ultra thick] (0.,0.) -- (-1.714,-1.);
\draw[blue, ultra thick] (0.,0.) -- (1.714,-1.);
\draw[red, decorate,decoration={snake,segment length=.12cm, amplitude=.04cm}] (-1.143,-0.667) -- (-1.643,0.199);
\node [centered, rotate=0.] at (0.2,2.) {\scriptsize $1$};
\node [centered, rotate=0.] at (-1.8,-1.5) {\scriptsize $\alpha_{+-}$};
\node [centered, rotate=0.] at (1.8,-1.5) {\scriptsize $\alpha_{-+}$};
\fill [red] (-1.643,0.199) circle (4pt);

\draw[->] (2.5,0.) -- (3.5,0.);

\draw[] (6.,0.) -- (6.,2.);
\draw[blue, ultra thick] (6.,0.) -- (4.286,-1.);
\draw[blue, ultra thick] (6.,0.) -- (7.714,-1.);
\draw[red, decorate,decoration={snake,segment length=.12cm, amplitude=.04cm}] (5.429,-0.333) -- (5.929,-1.199);
\node [centered, rotate=0.] at (6.2,2.) {\scriptsize $1$};
\node [centered, rotate=0.] at (4.2,-1.5) {\scriptsize $\alpha_{+-}$};
\node [centered, rotate=0.] at (7.8,-1.5) {\scriptsize $\alpha_{-+}$};
\fill [red] (5.929,-1.199) circle (4pt);

\draw[->] (8.5,0.) -- (9.5,0.);

\draw[] (12.,0.) -- (12.,2.);
\draw[blue, ultra thick] (12.,0.) -- (10.286,-1.);
\draw[blue, ultra thick] (12.,0.) -- (13.714,-1.);
\draw[red, decorate,decoration={snake,segment length=.12cm, amplitude=.04cm}] (13.143,-0.667) -- (12.643,-1.533);
\node [centered, rotate=0.] at (12.2,2.) {\scriptsize $1$};
\node [centered, rotate=0.] at (10.2,-1.5) {\scriptsize $\alpha_{+-}$};
\node [centered, rotate=0.] at (13.8,-1.5) {\scriptsize $\alpha_{-+}$};
\fill [red] (12.643,-1.533) circle (4pt);

\draw[->] (14.5,0.) -- (15.5,0.);

\draw[] (18.,0.) -- (18.,2.);
\draw[blue, ultra thick] (18.,0.) -- (16.286,-1.);
\draw[blue, ultra thick] (18.,0.) -- (19.714,-1.);
\draw[red, decorate,decoration={snake,segment length=.12cm, amplitude=.04cm}] (18.571,-0.333) -- (19.071,0.533);
\node [centered, rotate=0.] at (18.2,2.) {\scriptsize $1$};
\node [centered, rotate=0.] at (16.2,-1.5) {\scriptsize $\alpha_{+-}$};
\node [centered, rotate=0.] at (19.8,-1.5) {\scriptsize $\alpha_{-+}$};
\fill [red] (19.071,0.533) circle (4pt);

\draw[] (0.,-5.) -- (0.,-3.);
\draw[blue, ultra thick] (0.,-5.) -- (-1.714,-6.);
\draw[blue, ultra thick] (0.,-5.) -- (1.714,-6.);
\draw[red, decorate,decoration={snake,segment length=.12cm, amplitude=.04cm}] (0.,-4.333) -- (-1.,-4.333);
\node [centered, rotate=0.] at (0.2,-3.) {\scriptsize $1$};
\node [centered, rotate=0.] at (-1.8,-6.5) {\scriptsize $\alpha_{+-}$};
\node [centered, rotate=0.] at (1.8,-6.5) {\scriptsize $\alpha_{-+}$};
\fill [red] (-1.,-4.333) circle (4pt);

\draw[->] (2.5,-5.) -- (15.5,-5.);

\draw[] (18.,-5.) -- (18.,-3.);
\draw[blue, ultra thick] (18.,-5.) -- (16.286,-6.);
\draw[blue, ultra thick] (18.,-5.) -- (19.714,-6.);
\draw[red, decorate,decoration={snake,segment length=.12cm, amplitude=.04cm}] (18.,-3.667) -- (19.,-3.667);
\node [centered, rotate=0.] at (18.2,-3.) {\scriptsize $1$};
\node [centered, rotate=0.] at (16.2,-6.5) {\scriptsize $\alpha_{+-}$};
\node [centered, rotate=0.] at (19.8,-6.5) {\scriptsize $\alpha_{-+}$};
\fill [red] (19.,-3.667) circle (4pt);

\draw[] (8.8,-2.2) -- (8.8,-4.3);
\draw[] (9.2,-2.2) -- (9.2,-4.3);
\node [centered, rotate=0.] at (9.5,-3.2) {$?$};
\end{tikzpicture}
}

\newcommand{\HalfBraidingA}{
\begin{tikzpicture}[baseline={([yshift=-.8ex]current bounding box.center)},scale=1.]
\draw [->-] (0, 0) -- (0, .7);
\draw [->-] (.5, 0) -- (.5, .7);
\draw [] (-.2, .7) -- (.7, .7) -- (.7, 1.3) -- (-.2, 1.3) -- (-.2, .7);
\draw [->-] (0, 1.3) -- (0, 2.);
\draw [->-] (.5, 1.3) -- (.5, 2.);
\node [centered, rotate=0.] at (-.3, .2) {\scriptsize $X_J$};
\node [centered, rotate=0.] at (.7, .2) {\scriptsize $y$};
\node [centered, rotate=0.] at (-.2, 1.8) {\scriptsize $y$};
\node [centered, rotate=0.] at (.8, 1.8) {\scriptsize $X_J$};
\node [centered, rotate=0.] at (.25, 1.) {\scriptsize $c_{X_J, y}$};
\end{tikzpicture}\qquad =\qquad
\begin{tikzpicture}[baseline={([yshift=-.8ex]current bounding box.center)},scale=1.]
\draw [->-] (1, 0) -- (1, 1);
\draw [->-] (1, 1) -- (1, 2);
\draw [->-] (.5, 0) arc [start angle = 180, end angle = 100, x radius = .5, y radius = 1.];
\draw [-<-] (1.5, 2) arc [start angle = 0, end angle = -80, x radius = .5, y radius = 1.];
\node [left, rotate=0.] at (0.5, .25) {\scriptsize $X_J$};
\node [right, rotate=0.] at (1.5, 1.75) {\scriptsize $X_J$};
\node [centered, rotate=0.] at (1.2, .2) {\scriptsize $y$};
\node [centered, rotate=0.] at (.8, 1.8) {\scriptsize $y$};
\end{tikzpicture}
}

\newcommand{\HalfBraidingB}{
\begin{tikzpicture}[baseline={([yshift=-.8ex]current bounding box.center)},scale=.8]
\draw [->-] (1, 0) -- (1, 1);
\draw [->-] (1, 1) -- (1, 2);
\draw [->-] (.5, 0) arc [start angle = 180, end angle = 100, x radius = .5, y radius = 1.];
\draw [-<-] (1.5, 2) arc [start angle = 0, end angle = -80, x radius = .5, y radius = 1.];
\node [left, rotate=0.] at (0.5, .25) {\scriptsize $X_J$};
\node [right, rotate=0.] at (1.5, 1.75) {\scriptsize $X_J$};
\node [centered, rotate=0.] at (1.2, .2) {\scriptsize $y$};
\node [centered, rotate=0.] at (.8, 1.8) {\scriptsize $y$};
\end{tikzpicture}\ = \ \bigoplus_{p,q\in L_J}\ \bigoplus_{k\in L_\Fus}\ \sqrt{\frac{d_k}{d_y}}\ z_{pqy}^{J;k}\quad
\begin{tikzpicture}[baseline={([yshift=-.8ex]current bounding box.center)},scale=.8]
\draw [->-] (1, 0) -- (1, .8);
\draw [->-] (1, 1.2) -- (1, 2);
\draw [->-] (1, .8) -- (1, 1.2);
\draw [->-] (.5, 0) arc [start angle = 180, end angle = 90, x radius = .5, y radius = .7];
\draw [-<-] (1.5, 2) arc [start angle = 0, end angle = -90, x radius = .5, y radius = .7];
\node [left, rotate=0.] at (0.5, .25) {\scriptsize $p$};
\node [right, rotate=0.] at (1.5, 1.75) {\scriptsize $q$};
\node [centered, rotate=0.] at (1.2, .2) {\scriptsize $y$};
\node [centered, rotate=0.] at (.8, 1.8) {\scriptsize $y$};
\node [centered, rotate=0.] at (1.2, 1.) {\scriptsize $k$};
\end{tikzpicture}
}

\newcommand{\HalfBraidingC}{
\begin{tikzpicture}[baseline={([yshift=-.8ex]current bounding box.center)},scale=1]
\draw [->-] (1.15, .4) arc (-60:30:.5);
\draw [->-] (.6, .1) -- (.9, .3);
\draw [->-] (1.2, 1.3) -- (1., 1.6);

\draw [->-] (1, 0) -- (1, 1);
\draw [->-] (.2, 1.5) -- (1, 1);
\draw [->-] (1.8, 1.5) -- (1, 1);
\node [centered, rotate=0.] at (1.15, 0.) {\scriptsize $y$};
\node [centered, rotate=0.] at (.2, 1.3) {\scriptsize $u$};
\node [centered, rotate=0.] at (1.8, 1.3) {\scriptsize $v$};
\node [centered, rotate=0.] at (.4, 0.) {\scriptsize $p$};
\node [centered, rotate=0.] at (.8, 1.7) {\scriptsize $q$};
\end{tikzpicture}\qquad = \qquad 
\begin{tikzpicture}[baseline={([yshift=-.8ex]current bounding box.center)},scale=1]
\draw [->] (.5, .5) arc (-170:-240:1.);
\draw [->] (.5, .5) arc (-170:-190:1.);
\fill [white] (.55, 1.3) circle (.2);

\draw [->-] (1, 0) -- (1, 1);
\draw [->-] (.2, 1.5) -- (1, 1);
\draw [->-] (1.8, 1.5) -- (1, 1);
\node [centered, rotate=0.] at (1.15, 0.) {\scriptsize $y$};
\node [centered, rotate=0.] at (.2, 1.3) {\scriptsize $u$};
\node [centered, rotate=0.] at (1.8, 1.3) {\scriptsize $v$};
\node [centered, rotate=0.] at (.5, .3) {\scriptsize $p$};
\node [centered, rotate=0.] at (1.1, 1.6) {\scriptsize $q$};
\end{tikzpicture}
}

\newcommand{\HalfBraidingD}{
\begin{tikzpicture}[baseline={([yshift=-.8ex]current bounding box.center)},scale=.8]
\draw [->-] (1, 0) -- (1, 1);
\draw [->-] (1, 1) -- (1, 2);
\draw [->-] (.5, 0) arc [start angle = 180, end angle = 100, x radius = .5, y radius = 1.];
\draw [-<-] (1.5, 2) arc [start angle = 0, end angle = -80, x radius = .5, y radius = 1.];
\node [left, rotate=0.] at (0.5, .25) {\scriptsize $p$};
\node [right, rotate=0.] at (1.5, 1.75) {\scriptsize $q$};
\node [centered, rotate=0.] at (1.2, .2) {\scriptsize $j_E$};
\node [centered, rotate=0.] at (.8, 1.8) {\scriptsize $j_E$};
\end{tikzpicture} =
\sum_{k\in L_\Fus} \sqrt{\frac{d_k}{d_{j_E}}}\ {z_{pqj_E}^{J;k}} 
\begin{tikzpicture}[baseline={([yshift=-.8ex]current bounding box.center)},scale=.8]
\draw [->-] (1, 0) -- (1, .8);
\draw [->-] (1, 1.2) -- (1, 2);
\draw [->-] (1, .8) -- (1, 1.2);
\draw [->-] (.5, 0) arc [start angle = 180, end angle = 90, x radius = .5, y radius = .7];
\draw [-<-] (1.5, 2) arc [start angle = 0, end angle = -90, x radius = .5, y radius = .7];
\node [left, rotate=0.] at (0.5, .25) {\scriptsize $p$};
\node [right, rotate=0.] at (1.5, 1.75) {\scriptsize $q$};
\node [centered, rotate=0.] at (1.2, .2) {\scriptsize $j_E$};
\node [centered, rotate=0.] at (.8, 1.8) {\scriptsize $j_E$};
\node [centered, rotate=0.] at (1.2, 1.) {\scriptsize $k$};
\end{tikzpicture} =
W_E^{J;pq}\
\begin{tikzpicture}[baseline={([yshift=-.8ex]current bounding box.center)},scale=.8]
\draw [->-] (1, 0) -- (1, 2);
\node [centered, rotate=0.] at (1.3, 1.) {\scriptsize $j_E$};
\end{tikzpicture}\ .
}

\newcommand{\ExcitedA}{
\begin{tikzpicture}[baseline={([yshift=-.8ex]current bounding box.center)},scale=.7]
\draw [->-] (3.2,5.6) -- (3.2,7.2);
\node [centered, rotate=0.] at (2.9,6.4) {\scriptsize $j$};
\end{tikzpicture}
}

\newcommand{\ExcitedB}{
\begin{tikzpicture}[baseline={([yshift=-.8ex]current bounding box.center)},scale=.8]
\draw [->-] (3.2,5.6) -- (3.2,6.1);
\draw [->-] (3.2,6.1) -- (3.2,6.7);
\draw [->-] (3.2,6.7) -- (3.2,7.2);
\draw [decorate,decoration={snake,segment length=.12cm, amplitude=.04cm}, ->] (3.2,6.7) -- (4.,6.7);
\draw [decorate,decoration={snake,segment length=.12cm, amplitude=.04cm}, ->] (3.2,6.1) -- (2.4,6.1);
\node [centered, rotate=0.] at (3.,5.8) {\scriptsize $j$};
\node [centered, rotate=0.] at (3.,6.4) {\scriptsize $k$};
\node [centered, rotate=0.] at (3.4,7.) {\scriptsize $j$};
\node [centered, rotate=0.] at (4.2,6.7) {\scriptsize $q$};
\node [centered, rotate=0.] at (2.2,6.1) {\scriptsize $p^\ast$};
\end{tikzpicture}
}

\newcommand{\NonAbA}{
\begin{tikzpicture}[baseline={([yshift=-.8ex]current bounding box.center)},scale=.8]
\draw [->-] (3.2,5.6) -- (3.2,6.1);
\draw [->-] (3.2,6.1) -- (3.2,6.7);
\draw [->-] (3.2,6.7) -- (3.2,7.2);
\draw [->-] (3.2,7.2) -- (3.9,7.6);
\draw [->-] (2.5,7.6) -- (3.2,7.2);
\draw [->-] (2.5,5.2) -- (3.2,5.6);
\draw [->-] (3.2,5.6) -- (3.9,5.2);
\draw [decorate,decoration={snake,segment length=.12cm, amplitude=.04cm}, ->] (3.2,6.7) -- (4.,6.7);
\draw [decorate,decoration={snake,segment length=.12cm, amplitude=.04cm}, ->] (3.2,6.1) -- (2.4,6.1);
\node [centered, rotate=0.] at (2.4,5.) {\scriptsize $e_n$};
\node [centered, rotate=0.] at (4.1,5.) {\scriptsize $i_n$};
\node [centered, rotate=0.] at (4.1,7.8) {\scriptsize $i_2$};
\node [centered, rotate=0.] at (2.4,7.8) {\scriptsize $e_1$};
\node [centered, rotate=0.] at (3.,5.8) {\scriptsize $j$};
\node [centered, rotate=0.] at (3.,6.4) {\scriptsize $j$};
\node [centered, rotate=0.] at (3.4,7.) {\scriptsize $j$};
\node [centered, rotate=0.] at (4.4,6.7) {\scriptsize $1_{\mathcal{I}, 1}$};
\node [centered, rotate=0.] at (2.,6.1) {\scriptsize $1_{\mathcal{I},1}$};
\end{tikzpicture}
}

\newcommand{\NonAbB}{
\begin{tikzpicture}[baseline={([yshift=-.8ex]current bounding box.center)},scale=.8]
\draw [->-] (3.2,5.6) -- (3.2,6.1);
\draw [->-] (3.2,6.1) -- (3.2,6.7);
\draw [->-] (3.2,6.7) -- (3.2,7.2);
\draw [->-] (3.2,7.2) -- (3.9,7.6);
\draw [->-] (2.5,7.6) -- (3.2,7.2);
\draw [->-] (2.5,5.2) -- (3.2,5.6);
\draw [->-] (3.2,5.6) -- (3.9,5.2);
\draw [decorate,decoration={snake,segment length=.12cm, amplitude=.04cm}, ->] (3.2,6.7) -- (4.,6.7);
\draw [decorate,decoration={snake,segment length=.12cm, amplitude=.04cm}, ->] (3.2,6.1) -- (2.4,6.1);
\node [centered, rotate=0.] at (2.4,5.) {\scriptsize $e_n$};
\node [centered, rotate=0.] at (4.1,5.) {\scriptsize $i_n$};
\node [centered, rotate=0.] at (4.1,7.8) {\scriptsize $i_2$};
\node [centered, rotate=0.] at (2.4,7.8) {\scriptsize $e_1$};
\node [centered, rotate=0.] at (3.,5.8) {\scriptsize $j$};
\node [centered, rotate=0.] at (3.,6.4) {\scriptsize $k$};
\node [centered, rotate=0.] at (3.4,7.) {\scriptsize $j$};
\node [centered, rotate=0.] at (4.5,6.7) {\scriptsize $q_{J,\beta}$};
\node [centered, rotate=0.] at (1.9,6.) {\scriptsize $p^\ast_{J^\ast,\alpha}$};
\end{tikzpicture}
}

\begin{document}

\title{Nonlinear Symmetry-Fragmentation of Nonabelian Anyons In Symmetry-Enriched Topological Phases: A String-Net Model Realization}

\date{\today}
\author[a]{Nianrui Fu}
\author[a]{Siyuan Wang}
\author[a]{Yu Zhao\footnote{Corresponding author}}
\author[a,b]{Yidun Wan\footnote{Corresponding author}}
\affiliation[a]{State Key Laboratory of Surface Physics, Department of Physics, Center for Field Theory and Particle Physics, and Institute for Nanoelectronic devices and Quantum computing, Fudan University, Shanghai 200433, China}
\affiliation[b]{Hefei National Laboratory, Hefei 230088, China}
\emailAdd{nrfu25@m.fudan.edu.cn, siyuanwang18@fudan.edu.cn,  yuzhao20@fudan.edu.cn, ydwan@fudan.edu.cn}

\abstract{Symmetry-enriched topological (SET) phases combine intrinsic topological order with global symmetries, giving rise to novel symmetry phenomena. While SET phases with Abelian anyons are relatively well understood, those involving nonabelian anyons remain elusive. This obscurity stems from the multi-dimensional internal gauge spaces intrinsic to nonabelian anyons -- a feature first made explicit in \cite{Hu2018} and further explored in our recent works \cite{Wang2020,zhao2022,zhao2024,zhao2024a,zhao2024b,zhao2025c}. These internal spaces can transform in highly nontrivial ways under global symmetries. In this work, we employ an exactly solvable model -- the multifusion Hu–Geer–Wu string-net model introduced in a companion paper \cite{fu2025symmetryenrichedtopologicalphasesgauging} -- to reveal how the internal gauge spaces of nonabelian anyons transform under symmetries. We uncover a universal phenomenon, global symmetry fragmentation (GSF), whereby symmetry-invariant anyons exhibit internal Hilbert space decompositions into eigensubspaces labeled by generally fractional symmetry charges. Meanwhile, symmetry-permuted anyons hybridize and fragment their internal spaces in accordance with their symmetry behavior. These fragmented structures realize genuinely nonlinear symmetry representations that transcend conventional linear and projective classifications. Our results identify nonlinear global symmetry fragmentation as a hallmark of SETs and may shed new light on symmetry-enabled control in topological quantum computation.}

\maketitle

\flushbottom

\section{Introduction}\label{sec:intro}

Symmetry-enriched topological (SET) phases arise when topological phases are endowed with global symmetries\cite{Mesaros2011,Hung2013,HungWan2013a,Lu2013,Gu2014a,Barkeshli2014c,Chang2015,heinrich2016symmetry,cheng2017exactly,Williamson2017,williamson2017a,Ning2018,Hu2022,wang2022exactly}, yielding structures\cite{HungWan2013a,Chang2015,heinrich2016symmetry,cheng2017exactly,Williamson} believed to extend beyond both the Landau-Ginzburg paradigm\cite{Levin2004,Hung2012a,chen2013critical,wang2015field,lee2018}\footnote{A recent paper by two of us\cite{zhao2025c} shows that topological phases and their phase transitions can be encompassed by a generalized Landau-Ginzburg Paradigm.} and ordinary topological order. SET phases may be classified by symmetry fractionalization\cite{cheng2017exactly,yao2010symmetry,Essin2012,barkeshli2014coherent,barkeshli2019symmetry,song2015space,Ning2018,li2024}; however, symmetry fractionalizations in general do not directly correspond to how anyons are transformed under global symmetries\cite{cheng2017exactly}, especially for nonabelian anyons. Symmetry transformations of nonabelian anyons in SET phases remain elusive. The key point lies in the internal gauge spaces of nonabelian anyons: Unlike Abelian anyons, which are one-dimensional, nonabelian anyons may carry multiple flux types and gauge charge sectors\cite{Hu2018,zhao2024b,Sarma2015,Stern2008} -- and thus bearing multi-dimensional nontrivial internal Hilbert spaces and transforming under symmetry in sophisticated ways: not merely acquiring a phase factor. 

Hence, it is essential to elucidate the nonabelian anyons' internal spaces in order to describe their transformations under global symmetry actions. Nevertheless, conventional topological quantum field theory (TQFT) approaches treat anyons as abstract simple objects in modular tensor categories\cite{Lan2014b,kong2022invitationtopologicalorderscategory}, hiding their internal gauge structures. Our recently constructed enlarged Hu-Geer-Wu (HGW) string-net model of topological phases\cite{zhao2025c} enabled explicit investigation of nonabelian anyons' internal spaces and was further extended to a model of SET phases\cite{zhao2025c, fu2025symmetryenrichedtopologicalphasesgauging}. Using anyons' internal spaces, we will show remarkable symmetry structures of nonabelian topological orders.

\begin{figure}[!ht]
\centering
\Lattice
\caption{HGW string-net model lattice. Each plaquette has a tail (wavy line).}
\label{fig:lat}
\end{figure}

Although our approach is generic, to be concrete, we consider a specific example: the $S_3$ quantum‐double ($D(S_3)$) phase enriched by its electromagnetic‐exchange ($\mathbb{Z}_2$) symmetry. We find that the EM-exchange symmetry exchanges the $D(S_3)$ anyon types $C$ and $F$, while preserving other anyon types. More importantly, the symmetry transforms the $D(S_3)$ anyons' internal Hilbert spaces nontrivially: The internal Hilbert spaces of certain types of anyons may fragment into eigensubspaces labeled by distinct (and generally fractional) global symmetry charges, which characterizes genuinely nonlinear (beyond projective) representations of the EM-exchange symmetry, due to the symmetry's categorical nature. As for anyon types $C$ and $F$, their internal spaces are hybridized and then fragmented. We shall call fragmentation of anyons' internal spaces global symmetry fragmentation (GSF), which according to our findings is a ubiquitous phenomenon of symmetry-entriched non‑Abelian topological phases. Understanding GSF may also offer new insights and means of realizing universal topological quantum computation -- particularly given recent evidence that $D(S_3)$ model can support universal quantum computation\cite{lo2025} -- and suggests routes for engineering novel symmetry‑enriched quantum materials.

\section{\eqs{S_3} Quantum Double Phase in the \eqs{S_3} Enlarged HGW String-Net Model}\label{sec:stqd}

We now briefly introduce the $D(S_3)$ topological phase realized in the enlarged HGW string-net model\cite{zhao2025c} with input UFC $\Vec(S_3)$. This model is a lattice gauge theory coupled with anyonic matter defined on a \(2\)-dimensional trivalent spatial lattice. Each plaquette has a tail anchored to its perimeter\footnote{The original string-net model in Ref. \cite{Levin2004}, which has no such tails, cannot fully describe dyon excitations. These tails carry anyons' fluxes, thus enlarging the Hilbert space to encompass the complete anyon spectrum of the model.}. Each edge/tail is oriented and labeled by a group element in
\(S_3 = \langle r, s \mid r^3 = s^2 = (sr)^2 = 1\rangle\) -- the configuration space of the gauge field. Reversing an edge/tail while inversing its group element preserves the gauge field configuration. The gauge-field Hilbert space is spanned by all possible gauge-field configurations, obeying \(S_3\) group multiplication rules at each vertex: Let $i,j,k\in S_3$ be counterclockwise respectively on the three edges oriented toward a vertex, multiplication \(ijk = 1\in S_3\) must hold.

In the enlarged HGW string-net model, the \(S_{3}\) lattice gauge field couples to the eight anyon types of \(D(S_{3})\): \(A,B,C,D,E,F,G,H\), which are the matter fields residing in plaquettes. Because \(S_{3}\) is nonabelian, type $C,D,E,F,G$, and $H$ anyons are \emph{nonabelian}, each occupying a multi-dimensional internal gauge space\cite{Iguchi1997,Guruswamy1999a,Barkeshli2013,Hu2013,Li2018,Li2019a}, analogous to quark colour spaces. But standard TQFTs overlook the internal gauge dofs because they treat each anyon as a structureless simple object in a modular tensor category. For a certain type $J$ anyon in a plaquette, its internal basis states are labeled by a pair $(p, \alpha)$. Here, \(p\in S_3\) is the \textbf{flux type} that sources the gauge curvature in the plaquette via the magnetic Gauss law, which dictates a commensurate gauge-field configuration on the tail in the plaquette. A flux type $p$ may have an associated charge space, spanned by the label \(1\le\alpha\le{\tt dim}\text{(charge space)}\). In the current example, a charge space is a representation space of \(S_{3}\). Abelian anyons only have a unique flux type \(p\) and one-dimensional charge space, while a nonabelian anyon may admit several fluxes and/or multi-dimensional charge spaces. Such internal spaces transform covariantly under local gauge transformations \cite{zhao2025c}. Table \ref{tab:anyon species} records the 8 $D(S_3)$ anyon types, their fluxes, and charges.
\begin{table}[htb]
\centering
\begin{tabular}{|c||c|c|c|c|c|c|c|c|}
    \hline
    \textbf{Anyon} & $A$ & $B$ & $C$ & $D$ & $E$ & $F$ & $G$ & $H$ \\
    \hline
    \textbf{Flux} & $1$ & $1$ & $1$ & $s,rs,sr$ & $s,rs,sr$ & $r,r^2$ & $r,r^2$ & $r,r^2$ \\ \hline
    \textbf{Charge} & $1$ & $1$ & $1,2$ & $1$ & $1$ & $1$ & $1$ & $1$ \\ 
    \hline
\end{tabular}
\caption{Anyon species of $D(S_3)$ phase.}
\label{tab:anyon species}
\end{table}

In a pure topological phase, an anyon's internal states are not observable: They would transform as the anyon hops on the lattice and inevitably interacts with the gauge field. An anyon's only gauge-invariant quantity or physical observable is its type.

The $D(S_3)$ phase admits a $Z_2$ symmetry that exchanges its anyon types $C$ and $F$, while preserving all other anyon types (see Appendix \ref{sec:anyon}). This $\Z_2$ symmetry is an EM-exchange because the type-$C$ anyon has only a trivial flux but a 2-dimensional charge space, while the $F$-type anyon has two flux types but trivial charge space.

\section{Enriching the \eqs{S_3} Quantum Double with EM-Exchange Symmetry}\label{sec:set}

Here we describe how to extend the enlarged HGW model describing the $D(S_3)$ phase to a model describing the EM-exchange symmetry-enriched $D(S_3)$ phase. We focus on this particular example, while the general construction is reported in a companion paper\cite{fu2025symmetryenrichedtopologicalphasesgauging}.

The original Levin-Wen string-net model can describe SET phases if its input UFCs are multifusion categories\cite{heinrich2016symmetry, Chang2015, cheng2017exactly}; however, such models have two limitations. First, they cannot manifest anyons' internal spaces, thereby unable to investigate symmetry actions on nonabelian anyons properly. Second, they require a manual endowment of global symmetries on the topological phases rather than using only the input UFCs characterizing pure topological phases. Our SET model overcomes these two limitations because (1) it is based on the enlarged HGW model that manifests anyons' internal spaces, and (2) it promotes the input UFCs of the enlarged HGW model to multifusion categories without \textit{ad hoc} knowledge of the global symmetries to be imposed. 

In our companion work\cite{fu2025symmetryenrichedtopologicalphasesgauging}, we can first derive multifusion categories, each equipped with an appropriate isomorphism (necessary for defining symmetry sectors), from the input UFC $\Fus$ of the enlarged HGW model. Then, by replacing $\Fus$ with one of these derived multifusion category together with a chosen isomorphism as the input data of the enlarged HGW model, we obtain an exactly solvable lattice model of an SET phase. This model is reviewed in Appendix \ref{sec:review}. 

Such a multifusion category is derived from the irreducible bimodules over certain Frobenius algebra objects of $\Fus$ \cite{fu2025symmetryenrichedtopologicalphasesgauging}. Here, we focus on the case where $\Fus = \Vec(S_3)$. For the model describing the EM-exchange symmetry-enriched $D(S_3)$ phase, we directly adopt from Ref. \cite{fu2025symmetryenrichedtopologicalphasesgauging} the corresponding input multifusion category equipped with the necessary isomorphism derived from $\Vec(S_3)$. This multifusion category can take a $2\times 2$ matrix form:
\begin{equation}\label{eq:s3multifusion}
    \mathcal{M}=\left(\begin{array}{cc}
        \{1_{++},r_{++},r^2_{++},s_{++},rs_{++},sr_{++}\} & \{\alpha_{+-},\ \beta_{+-}\} \\
        \{\alpha_{-+},\ \beta_{-+}\} & \{1_{--},r_{--},r^2_{--},s_{--},rs_{--},sr_{--}\}
    \end{array}\right).
\end{equation}

The two diagonal sets of elements of the multi-fusion category $\mathcal{M}$, i.e., $\{1_{++},\allowbreak r_{++},\allowbreak r^2_{++},\allowbreak s_{++},\allowbreak rs_{++},\allowbreak sr_{++}\}$ and $\{1_{--},\allowbreak r_{--},\allowbreak r^2_{--},\allowbreak s_{--},\allowbreak rs_{--},\allowbreak sr_{--}\}$, are two isomorphic UFCs and both isomorphic to $\Vec(S_3)$. They are respectively the sets of the basic degrees of freedom of the two different symmetry sectors, labled by $+$ and $-$, related by the EM-exchange symmetry. The off-diagonal sets in $\mathcal{M}$ comprise the domain-wall degrees of freedom defined as follows. A domain wall between symmetry sectors $+$ and $-$ is oriented: Along its orientation, a domain wall's left (right) always lies in the $+$ ($-$) sector, so its degree of freedom takes value in the set $\{\alpha_{+-}, \beta_{+-}\}$. If we reverse the orientation of a domain wall, say, $\alpha_{+-}$, it should become $\alpha_{-+}$, and its left and right sectors are exchanged (see Eq. \ref{eq:dw}). 
\begin{equation}\label{eq:dw}
    \DW
\end{equation}
The EM-exchange symmetry transformation $\mathcal{G}$ swaps the two sectors:
\begin{equation}
    \mathcal{G}:a_{ij}\to a_{\bar{i}\bar{j}},\ \forall a\in\{1,r,r^2,s,rs,sr\},\ i,j\in\{+,-\},
\end{equation}
where, $\bar{+}=-,\ \bar{-}=+$. 

The SET model still has anyon excitations, which can be extracted from the half braiding $z$-tensor of the multifusion category \eqref{eq:s3multifusion} (see Appendix \ref{sec:categorical_data}). More importantly, we can read off directly from the $z$-tensors how the $D(S_3)$ anyons are transformed under the symmetry. For example, when a type-$C$ anyon in the $+$ sector hops into the $-$ sector, it becomes a type-$F$ anyon. Let's elaborate.

\begin{figure}
    \centering
    \SymmTrans
    \caption{Anyon $a$ crossing a domain wall (thick line) is transformed to be anyon $\mathcal{G}(a)$.}
    \label{fig:symmtrans}
\end{figure}

\section{Global Symmetry Fragmentation}

As different symmetry sectors differ by a symmetry transformation, when an anyon in one symmetry sector crosses a domain wall and enters another symmetry sector, it undergoes a symmetry transformation, conducted by the domain wall, as depicted in Figure \ref{fig:symmtrans}. For example, a $C$ (an $F$) anyon in sector $+$ upon crossing a domain wall would be transformed into an $F$ (a $C$) anyon in sector $-$. The conventional methods of studying SET phases would have to stop at this level of understanding of the symmetry transformations of nonabelian anyons. Nonetheless, by means of our domain wall constructed in the enlarged HGW model, explicit transformations on anyons' internal spaces can be revealed. 

As aforementioned, an anyon's internal states change when the anyon hops on the lattice. So, to focus only on global symmetry transformations on an anyon's internal space, we should consider the scenario where the anyon sits in a plaquette right beside a domain wall, then crosses the domain wall into the plaquette on the other side. 

Back to the EM-exchange symmetry enriched $D(S_3)$ phase. Although according to Eq. \eqref{eq:s3multifusion}, apparently there are two types of domain walls labeled by $\alpha$ and $\beta$, they in fact differ by a $\Z_2$ gauge generated by the gauge degree of freedom $s\in S_3$ with $s^2=1$.\cite{fu2025symmetryenrichedtopologicalphasesgauging} As such, if we know the symmetry transformations on the anyons conducted by crossing domain wall $\alpha$, we can easily derive those conducted by crossing domain wall $\beta$. Hence, hereafter, we would consider domain wall $\alpha$ only unless otherwise mentioned. The EM-exchange symmetry transformations on the $D(S_3)$ anyons can be directly read off from the $z$-tensors (see Appendix \ref{sec:categorical_data}) of the SET model. For instance, from $z^{(F,C);\alpha_{+-}}_{r_{++},(1_1)_{--},\alpha_{+-}}=1$, we can see that EM exchange transformation $\mathcal{G}_{em}$ turns $F_r$ into $C_{1_1}$. Table \ref{tab:transformation} records all such transformations, which give the representations $\rho^{\tilde J}(\mathcal{G}_{em})$ of the transformation $\mathcal{G}_{em}$ over the internal space of $\tilde J$, which is possibly a composite anyon, such as $C\oplus F$. For any $\tilde J$ the representation of trivial action $1$ is trivial: $\rho^{\tilde J}(1)=\idm$. 
For $\tilde J=D$ as an example, with $\omega:=e^{i\frac{2\pi}{3}}$:
\begin{equation}\label{eq:rhoD}
    \rho^D(\mathcal{G}_{em})=\frac{1}{\sqrt{3}}\begin{pmatrix}
        1 & 1 & 1 \\
        1 & \omega & \omega^* \\
        1 & \omega^* & \omega \\
    \end{pmatrix}.
\end{equation}

\begin{table}[h]
    \centering 
    \renewcommand{\arraystretch}{1.5}
    \begin{tabular}{|>{\centering\arraybackslash}m{3cm}|>{\centering\arraybackslash}m{5cm}|}
    \hline
        \textbf{Anyon Internal Basis} & \textbf{EM-exchange transformed basis} \\
        \hline
        $C_{1_1}$ & $F_r$  \\
        \hline
        $C_{1_2}$ & $F_{r^2}$  \\
        \hline
        $F_r$ & $C_{1_1}$  \\
        \hline
        $F_{r^2}$ & $C_{1_2}$  \\
        \hline
        $G_r$ & $\omega G_r$  \\
        \hline
        $G_{r^2}$ & $\omega G_{r^2}$  \\
        \hline
        $H_r$ & $\omega^*H_{r^2}$  \\
        \hline
        $H_{r^2}$ & $\omega^*H_{r}$  \\
        \hline
        $D_{s}$ & $(D_{s}+D_{rs}+D_{sr})/\sqrt{3}$  \\
        \hline
        $D_{rs}$ & $(D_{s}+\omega D_{rs}+\omega^*D_{sr})/\sqrt{3}$  \\
        \hline
        $D_{sr}$ & $(D_{s}+\omega^*D_{rs}+\omega D_{sr})/\sqrt{3}$  \\
        \hline
        $E_{s}$ & $(E_{s}+E_{rs}+E_{sr})/\sqrt{3}$  \\
        \hline
        $E_{rs}$ & $(E_{s}+\omega E_{rs}+\omega^*E_{sr})/\sqrt{3}$  \\
        \hline
        $E_{sr}$ & $(E_{s}+\omega^*E_{rs}+\omega E_{sr})/\sqrt{3}$  \\
        \hline
    \end{tabular}
    \caption{Nontrivial EM-exchange symmetry transformations on the internal states of $D(S_3)$ anyons, as conducted by hopping the anyons across an $\alpha$-type domain wall. Here, $J_{x}$ denotes the internal basis state $x$ of a $J$-anyon.}
    \label{tab:transformation}
\end{table}

We then have three observations: 
\begin{enumerate}

\item Certain nonabelian anyons' internal spaces are fragmented into subspaces as eigenspaces of the symmetry action. For example, the internal space of anyon $H$ is fragmented into two symmetry eigenstates: $H_{r}+H_{r^2}$ and $H_{r}-H_{r^2}$, with definite symmetry charges $\frac{2}{3}$ and $\frac{1}{6}$. (We will explain the nature of these charges shortly.) We dub this phenomenon on nonabelian anyons \emph{global symmetry fragmentation}, which is ubiquitous in any symmetry-enriched nonabelian topological phases. An exception is the $G$ type, whose internal space is not fragmented; rather, a $G$-type anyon as a whole acquires a global symmetry charge of $1/3$. The fragmentation of $D$ and $E$ anyons are somewhat more involved but can be easily derived from the $z$-tensors in Appendix \ref{sec:categorical_data}. It's worth of note that the symmetry fragmentation pattern in an SET phase is gauge dependent. As we mentioned earlier, the EM-exchange symmetry transformations can be defined as conducted by the $\beta$-type domain wall (called the $\beta$-gauge), which is related to those in Table \ref{tab:transformation} by the $\alpha$-type domain wall (the $\alpha$-gauge). If we take the $\beta$-gauge, it would be the $G$ but not $H$ anyons acquire fragmentation.  
\item Anyon types $C$ and $F$ are exchanged, and their internal spaces are mixed up: $C_{1_1}\pm F_r$ and $C_{1_2}\pm F_{r^2}$. So, the mixed internal space of $C$ and $F$ is fragmented into two 2-dimensional eigenspaces of the symmetry: 
\begin{equation}
    {\tt span}\{C_{1_1}\pm F_r,\ C_{1_2}\pm F_{r^2}\},
\end{equation}
respectively with charge $0$ and charge $1/2$.
\end{enumerate} 

\section{Representation of EM-Exchange Symmetry Is Nonlinear}

Do anyons' fractional symmetry charges label certain linear or projective representations of the symmetry group? Not necessary! In the current example, these charges do not correspond to any linear or projective representations of the EM-exchange symmetry because the symmetry group is $\Z_2$, which has linear representations labeled by $\pm 1$ only and has no projective representations at all. In fact, the fractional charges we have found label nonlinear representations of $\Z_2$. This can be shown by composing two consecutive symmetry transformations on the $D(S_3)$ anyons as follows. 

Since the symmetry transformations are conducted by hopping anyons cross a domain wall, hopping an anyon cross $2$ neighboring domain walls composes two such transformations on the anyon, as seen in Figure \ref{fig:symmcomposition}. Pachner moves\footnote{Pachner moves are topological moves. See Appendix \ref{sec:pachner}} are invoked in composing the two hoppings; accordingly, composing the corresponding two symmetry formations must involve the effects of the Pachner moves and thus cannot be directly multiplying the two representation matrices\footnote{Note that nonlinear representations are not matrices; however, here we assume a section in the representation bundle is chosen for us to express the representation as matrices. The nonlinearity is manifest in composing two symmetry transformations.} of the two transformations.

\begin{figure}[htb]
    \centering
    \Composition
    \caption{Composing symmetry transformations is equivalent to anyon crossing two domain walls (thick lines). Edge orientation is upward.}
    \label{fig:symmcomposition}
\end{figure}

In our EM-exchange symmetry enriched $D(S_3)$ phase, composing two nontrivial symmetry transformations must result in the trivial transformation, i.e.,  $\mathcal{G}^{\Tilde{J}}_{em}\circ\mathcal{G}^{\Tilde{J}}_{em} \equiv \idm $, which reads in the matrix form as
\begin{equation}\label{eq:composition}
    \sum_{b}\omega_{ab}\rho^{\Tilde{J}}_{ab}(\mathcal{G}_{em})\rho^{\Tilde{J}}_{bc}(\mathcal{G}_{em})=\rho^{\Tilde{J}}_{ac}(1)=\idm,
\end{equation}
where matrices $\rho^{\Tilde{J}}(\mathcal{G}_{em})$, e.g., \eqref{eq:rhoD}, can be extracted from Table \ref{tab:transformation}. Here, $\omega$ -- a rank-$2$ tensor due to the Pachner moves and independent of $\Tilde{J}$ -- has non-zero elements:
\begin{equation}
    \begin{split}
        &\omega_{1,r}=\omega_{1,r^2}=\omega_{1,s}=\omega_{1,rs}=\omega_{1,sr}=1,\\
        &\omega_{s,r}=\omega_{s,r^2}=\omega_{s,rs}=\omega_{s,sr}=1,\\
        &\omega_{r,r}=\omega_{r^2,r^2}=\omega_{rs,rs}=\omega_{sr,sr}=e^{i\frac{2\pi}{3}},\\
        &\omega_{r,r^2}=\omega_{rs,sr}=e^{-i\frac{2\pi}{3}}.\\
    \end{split}
\end{equation}
As opposed to the case of projective representations, $\omega$ is not a phase factor associated with an entire matrix $\rho^{\Tilde{J}}(\mathcal{G}_{em})$ but has different components for different matrix elements of $\rho^{\Tilde{J}}(\mathcal{G}_{em})$. Therefore, the representations $\rho^{\Tilde{J}}$ are neither linear nor projective, but nonlinear representations of the EM-exchange symmetry. The fractional charges carried by the fragmented internal spaces of $D(S_3)$ anyons precisely label the irreducible nonlinear representations. We demonstrate this nonlinearity in the following three instances from Table \ref{tab:transformation}.

\begin{enumerate}
    \item For $\Tilde{J}=H$, the basis states $H_r$ and $H_{r^2}$ are exchanged under GEM. Hence, the representation $\rho^{H}$ can be decomposed into two $1$-dimensional representations $\rho^{H_+}$ and $\rho^{H_-}$, where $H_+=H_r+H_{r^2}, H_-=H_r-H_{r^2}$. We have $\rho^{H_+}(\mathcal{G}_{em})=e^{-i\frac{2\pi}{3}},\ \rho^{H_-}(\mathcal{G}_{em})=e^{i\frac{\pi}{3}}$, such that \eqref{eq:composition} becomes
    \begin{equation}\label{eq:Hrep}
        \begin{aligned}
            & \rho^{H_+}(\mathcal{G}_{em})\rho^{H_+}(\mathcal{G}_{em})=e^{i\frac{2\pi}{3}}\rho^{H_+}(1);\\
            & \rho^{H_-}(\mathcal{G}_{em})\rho^{H_-}(\mathcal{G}_{em})=e^{i\frac{2\pi}{3}}\rho^{H_-}(1).
        \end{aligned}
    \end{equation}
    One might then mistake the irreducible representations $\rho^{H_+}$ and $\rho^{H_-}$ as projective representations of $\Z_2$ and further regard them as linear representations because any apparent projective representation of  $\mathbb{Z}_2$ is equivalent to a linear representation because $H^2(\mathbb{Z}_2, U(1)) = \{0\}$.  But we can show that $\rho^{H_+}$ and $\rho^{H_-}$ can never be made equivalent to linear representations of $\Z_2$ (see Appendix \ref{app:proof}). Thus, $H$ is indeed fragmented into two nontrivial nonlinear representations of $\Z_2$ characterized by charges $2/3$ and $1/6$, revealing the true meaning of these charges. 
    \item If $\Tilde{J}=G$, the two basis states $G_r$ and $G_r^2$ both acquire a phase factor $e^{i\frac{2\pi}{3}}$ under the symmetry action. So, \eqref{eq:composition} now reads
    \begin{equation}\label{eq:Grep}
        \begin{aligned}
            & \rho^{G_r}(\mathcal{G}_{em})\rho^{G_r}(\mathcal{G}_{em})=e^{-i\frac{2\pi}{3}}\rho^{G_r}(1);\\
            & \rho^{G_{r^2}}(\mathcal{G}_{em})\rho^{G_{r^2}}(\mathcal{G}_{em})=e^{-i\frac{2\pi}{3}}\rho^{G_{r^2}}(1).
        \end{aligned}
    \end{equation}
    Likewise in Appendix \ref{app:proof}, we can show that the composition of symmetry transformations above indeed signifies a nonlinear representation of $\Z_2$. One might however misconceive that $G$'s internal space is also fragmented to two copies of the same 1-dimensional nonlinear representation. In fact, as the basis states $G_r$ and $G_{r^2}$ are always interchangeable under $G$'s internal gauge, they are not physically distinguishable. Therefore, one should regard the 2-dimensional internal space of $G$ as an indecomposable 2-dimensional nonlinear representation space of $\Z_2$ characterized by charge $1/3$. 
    \item For $\Tilde{J} = C\oplus F$,  the 4-dimensional representation $\rho^{C\oplus F}$ could be decomposed into two $2$-dimensional representations $\rho^{(C\oplus F)_+}$ and $\rho^{(C\oplus F)_-}$, where
    \begin{equation}
    V_{(C\oplus F)_{\pm}} = {\tt span}\{C_{1_1}\pm F_r,\ C_{1_2}\pm F_{r^2}\},
    \end{equation} 
    such that $\rho^{(C\oplus F)_+}(\mathcal{G}_{em})=\idm$ and $\rho^{(C\oplus F)_-}(\mathcal{G}_{em})=-\idm$. These two 2-dimensional representations are irreducible for the following reason. Since the internal gauges of $C$ ($F$) can mix $C_{1_1}$ ($F_r$) and $C_{1_2}$ ($F_{r^2}$), we cannot differentiate $C_{1_1}+F_r$ ($C_{1_1}-F_r$) from $C_{1_2}+F_{r^2}$ ($C_{1_2}-F_{r^2}$). As such, all states in the 2-dimensional space $V_{(C\oplus F )_+}$ ($V_{(C\oplus F )_-}$) carry the same symmetry charge $0$ ($1/2$), rendering the space a degenerate eigenspace of the symmetry action. The composition rule \eqref{eq:composition} now takes the form
    \begin{equation}
        \rho^{(C\oplus F)_{\pm}}(\mathcal{G}_{em})\rho^{(C\oplus F)_{\pm}}(\mathcal{G}_{em})=\rho^{(C\oplus F)_{\pm}}(1).
    \end{equation}
\end{enumerate}

Therefore, it is clear that when the $D(S_3)$ phase is endowed with the EM-exchange symmetry, certain types of its nonabelian anyons do fragment into nonlinear representations under the symmetry. This justifies the terminology \emph{nonlinear symmetry Fragmentation} of nonabelian anyons. By the systematic construction of lattice models of SET phases in our companion paper\cite{fu2025symmetryenrichedtopologicalphasesgauging},  nonlinear symmetry fragmentation is a general feature of symmetry-enriched nonabelian topological phases. 

\section{Discussion}

In this paper, for the first time in the literature to date, we have explicitly studied the global symmetry action on nonabelian anyons in an SET phase -- the $D(S_3)$ phase enriched by the EM-exchange symmetry. We discovered the phenomenon of nonlinear symmetry fragmentation by showcasing how $D(S_3)$ nonabelian anyons's internal spaces are rearranged and decomposed into nonlinear irreducible representations of the EM-exchange (a $\Z_2$) symmetry. This phenomenon is not unique to this example but ubiquitous in SET phases involving nonabelian topological orders. A recent work\cite{lo2025} shows that the $D(S_3)$ nonabelian anyons can support universal quantum computation. Since now we have a finer understanding of how such anyons behave in a richer phase -- the $D(S_3)$ phase enriched by the EM-exchange symmetry, one may further explore the utility of nonlinear symmetry fragmentation for topological quantum computation, such as improving gate and/or algorithm efficiency by harnessing the power of global symmetries.

It would be an illuminating future work to study the interplay between nonabelian anyons and nonabelian global symmetry, which would arise in for example the $D_4$ quantum double phase enriched by the permutation symmetry $S_3$. Even more fascinating to study is: If a nonabelian topological phase is endowed with not a group but an an algebraic global symmetry, how would the nonabelian anyons therein transform under the global symmetry? How would the concept of nonlinear symmetry fragmentation be generalised in such cases?

\begin{acknowledgments}
The authors thank Hongguang Liu, Yuting Hu, and Yifei Wang for inspiring and helpful discussions. YW is supported by NSFC Grant No. KRH1512711, the Shanghai Municipal Science and Technology Major Project (Grant No. 2019SHZDZX01), Science and Technology Commission of Shanghai Municipality (Grant No. 24LZ1400100), and the Innovation Program for Quantum Science and Technology (No. 2024ZD0300101). YW is grateful for the hospitality of the Perimeter Institute during his visit, where the main part of this work is done. This research was supported in part by the Perimeter Institute for Theoretical Physics. Research at Perimeter Institute is supported by the Government of Canada through the Department of Innovation, Science and Economic Development and by the Province of Ontario through the Ministry of Research, Innovation and Science. 
\end{acknowledgments}

\appendix

\section{String-net Model}\label{sec:review}

In this section, we briefly review the string-net model defined in Ref. \cite{zhao2024}, which was adapted from that in \cite{Hu2018}. The string-net model is an exactly solvable model defined on a \(2\)-dimensional lattice. An example lattice is depicted in Fig. \ref{fig:lat}. All vertices are trivalent. Within each plaquette of the lattice, a tail is attached to an arbitrary edge of the plaquette, pointing inward. We will demonstrate that different choices of the edge to which the tail is attached are equivalent in Appendix \ref{sec:pachner}. Each edge and tail is oriented, but we'll show that different choices of directions are equivalent.


The input data of the string-net model is a unitary fusion category \(\Fus\), described by a finite set \(L_\Fus\), whose elements are called \emph{simple objects}, equipped with three functions \(N: L_\Fus^3 \to \NN\), \(d: L_\Fus \to \RR^+\), and \(G: L^6_\Fus \to \C\). The function \(N\) sets the \emph{fusion rules} of the simple objects, satisfying
\eq{
\sum_{e \in L_\Fus} N_{ab}^e N_{ec}^d = \sum_{f \in L_\Fus} N_{af}^d N_{bc}^f, \qquad\qquad N_{ab}^c = N_{c^\ast a}^{b^\ast}.
}
There exists a special simple object \(1 \in L_\Fus\), called the \emph{trivial object}, such that for any \(a, b \in L_\Fus\),
\eq{
N_{1a}^b = N_{1b}^a = \delta_{ab},
}
where \(\delta\) is the Kronecker symbol. For each \(a \in L_\Fus\), there exists a unique simple object \(a^\ast \in L_\Fus\), called the \emph{opposite object} of \(a\), such that
\eq{
N_{ab}^1 = N_{ba}^1 = \delta_{ba^\ast}.
}
We only consider the case where for any \(a, b, c \in L_\Fus\), \(N_{ab}^c = 0\) or \(1\). In this case, we define
\eq{
\delta_{abc} = N_{ab}^{c^\ast} \in \{0, 1\}.
}

The basic configuration of the string-net model is established by labeling each edge and tail with a simple object in \(L_\Fus\), subject to the constraint on all vertices that \(\delta_{ijk} = 1\) for the three incident edges or tails meeting at this vertex, all pointing toward the vertex and respectively counterclockwise labeled by \(i, j, k \in L_\Fus\). We can reverse the direction of any edge or tail and simultaneously conjugate its label as \(j \to j^\ast\), which keeps the configuration invariant. The Hilbert space \(\Hil\) of the model is spanned by all possible configurations of these labels on the edges and tails.

The function \(d: L_\Fus\to\RR^+\) returns the \emph{quantum dimensions} of the simple objects in \(L_\Fus\). It is the largest eigenvalues of the fusion matrix and forms the \(1\)-dimensional representation of the fusion rule.
\eq{d_ad_b = \sum_{c\in L_\Fus}N_{ab}^cd_c.}
In particular, \(d_1 = 1\), and for any \(a\in L_\Fus, d_a = d_{a^\ast}\ge 1\). 

The function \(G: L_\Fus^6\to\CC\) defines the \(6j\)-\emph{symbols} of the fusion algebra. It satisfies
\eqn[eq:sixj]{\sum_nd_nG^{pqn}_{v^*u^*a}G^{uvn}_{j^*i^*b}G^{ijn}_{q^*p^*c} = G^{abc}_{i^*pu^*}&G^{c^*b^*a^*}_{vq^*j},\qquad\sum_nd_nG^{ijp}_{kln}G^{j^*i^*q}_{l^*k^*n} = \frac{\delta_{pq^*}}{d_p}\delta_{ijp}\delta_{klq},\\
G^{ijm}_{kln} = G^{klm^*}_{ijn^*} = G^{jim}_{lkn^*}= G^{mij}_{nk^*l^*} &= \alpha_m\alpha_n\overline{G^{j^*i^*m^*}_{l^*k^*n^*}},\qquad \Big|G^{abc}_{1bc}\Big| = \frac{1}{\sqrt{d_bd_c}}\delta_{abc},
}
where \(\alpha_m = G^{1mm^\ast}_{1m^\ast m}\in\{\pm 1\}\) is the Frobenius-Schur indicator of simple object \(m\).

The Hamiltonian of the string-net model reads
\eqn{H := - \sum_{{\rm Plaquettes}\ P}Q_P,\qquad\qquad Q_P := \frac{1}{D}\sum_{s\in L_\Fus}d_sQ_P^s,}
where operator \(Q_P^s\) acts on edges surrounding plaquette \(P\) and has the following matrix elements on a hexagonal plaquette:
\begin{align*}Q_P^s\ \PlaquetteSrc\
:=\ \delta_{p,1}\delta_{j_1,j_7}\ \sum_{j_k\in L_\Fus}\ \prod_{k = 1}^{6}\ \Bigg(\sqrt{d_{i_k}d_{j_k}}G^{e_ki_ki_{k+1}^\ast}_{sj_{k+1}^\ast j_k}\Bigg)\PlaquetteTar\ ,
\end{align*}
and
\eq{D := \sum_{a\in L_\Fus}d_a^2}
is the total quantum dimension of UFC \(\Fus\). Here, we only show the actions of the \(Q_P\) operator on a hexagonal plaquette. The matrix elements of \(Q_P\) operators on other types of plaquettes are defined similarly. We also omit the ``$\ket{\cdot}$'' labels surrounding all diagram for simplicity, unless they are specifically required.

It turns out that
\eq{
(Q_P^s)^\dagger = Q_P^{s^\ast},\qquad Q_P^rQ_P^s = \sum_{t\in L_\Fus} N_{rs}^tQ_P^t,\qquad Q_P^2 = Q_P,\qquad Q_{P_1}Q_{P_2} = Q_{P_2}Q_{P_1}.
}
The summands \(Q_P\) in Hamiltonian \(H\) are commuting projectors, so the Hamiltonian is exactly solvable. The ground-state subspace \(\Hil_0\) of the system is the projection
\eqn{
\Hil_0 = \Bigg[\prod_{{\rm Plaquettes\ } P}Q_P\Bigg]\Hil.
}
If the lattice has the sphere topology, the model has a unique ground state \(\ket\Phi\) up to scalar factors.

\subsection{Topological Features}\label{sec:pachner}

We briefly review the topological nature of the ground-state subspace of the string-net model defined in Ref. \cite{Hu2018}. Topologically, any two lattices with the same topology can be transformed into each other by the \emph{Pachner moves}. There are unitary linear maps between the Hilbert spaces of two string-net models with the same input UFC on different lattices related by the Pachner moves\cite{Hu2012}, formally denoted as operators \(\T\). The ground states are invariant under such linear transformations. There are three kinds of elementary Pachner moves, whose corresponding linear transformations are:
\eqn[eq:pachner]{
&\T \quad \PachnerOne\ ,\\
&\T \quad \PachnerTwo\ ,\\
&\T \quad \PachnerThree\ .}
Here, ``\({\color{red}\times}\)'' marks a plaquette to be contracted. These
three elementary Pachner moves and their corresponding unitary transformations can be composed. Given initial and final lattices, there are multiple ways to compose these elementary Pachner moves, but different ways result in the same transformation matrices on the ground-state Hilbert space.

We have also noted that for a plaquette, different choices of edge to which the tail is attached are equivalent. These variations lead to distinct lattice configurations and, consequently, different Hilbert spaces for the lattice model. The equivalence between such two Hilbert spaces is established by the following linear transformation \(\T'\):
\eqn[eq:tailmove]{
\T'\quad\PachnerFour\ .
}
The states where tails attach to other edges can be obtained recursively in this manner.

For convenience, in certain cases, we will temporarily incorporate auxiliary states with multiple tails within a single plaquette. These states, despite having multiple tails in one plaquette, are all equivalent to states within the Hilbert space (with only one tail in each plaquette):
\eqn[eq:tailfuse]{
\PachnerFive\ .
}

\subsection{Excited States}\label{sec:spec}

An \emph{excited state} \(\ket\varphi\) of the string-net model is an eigenstate such that \(Q_P\ket\varphi = 0\) at some plaquettes \(P\). In such a state, we say there are \emph{anyons} in these plaquettes \(P\). We also refer to the ground states as trivial excited states, in which there are only \emph{trivial anyons} in all plaquettes. We assume the sphere topology, in which the model has a unique ground state; nevertheless, the results in this section apply to other topologies.

We start with the simplest excited states with a pair of anyons in two \emph{adjacent} plaquettes with a common edge \(E\). This state can be generated by ribbon operator \(W_E^{J;pq}\), which is a composition of an auxiliary operator \(\mathcal{W}_E^{J; pq}\)
\eqn[eq:create]{
\mathcal{W}_E^{J; pq} \ExcitedA := \sum_{k \in L_\Fus} \sqrt{\frac{d_k}{d_{j_E}}} \ {z_{pqj_E}^{J;k}} \ \ \ExcitedB \ 
}
and Pachner moves \eqref{eq:tailmove} and \eqref{eq:tailfuse}, where $j_E$ is the dof on edge $E$. For any state in the Hilbert space of the model, each plaquette contains exactly one tail attached to a specific edge (could be edge $E$) of the plaquette. Operator \(\mathcal{W}_E^{J; pq}\) \eqref{eq:create} introduces two new tails on edge $E$, and these new tails must be moved and fused with the original tails of the plaquettes by the subsequent Pachner moves in the creation opeartor.

The coefficients in Eq. \eqref{eq:create},  \(z_{pqj}^{J; k}\), are called the \emph{half-braiding tensor} of anyon type \(J\), defined by the following equation:
\eqn[eq:halfA]{
\frac{\delta_{jt}N_{rs}^t}{d_t} z_{pqt}^{J;w} = \sum_{u,l,v\in L_\Fus} z_{lqr}^{J;v} z_{pls}^{J;u}d_u d_v G^{r^*s^*t}_{p^*wu^*} G^{srj^*}_{qw^*v} G^{s^*ul^*}_{rv^*w}.
}
We will discuss the physical significance of this equation in Appendix \ref{sec:halfbraid}. The label \(J\), called the \emph{anyon type}, labels different minimal solutions \(z^J\) of Eq. \eqref{eq:halfA} that cannot be the sum of any other nonzero solutions. Categorically, anyon type \(J\) are labeled by simple objects in the \emph{center} of UFC \(\Fus\), a modular tensor category whose categorical data record all topological properties of the topological order that the string-net model describes, denoted as \(\Cent(\Fus)\). In particular, the trivial anyon $\mathcal{I}\in L_{\Cent(\Fus)}$ satisfies
$$z^{\mathcal{I};k}_{pqj} = \delta_{p1}\delta_{q1}\delta_{jk}.$$

The statistics of anyon \(J\) are recorded by the diagonal element of modular \(T\) matrix of UMTC \(\Cent(\Fus)\), where
\eq{
T_{JK} = \frac{1}{d_t}\delta_{JK}\sum_{p \in L_\Fus} d_p z_{ttt}^{J;p}.}
Here, \(t\) is an arbitrary charge of anyon \(J\). The braiding of two anyons \(J\) and \(K\) is recorded by the modular \(S\) matrix, whose matrix elements are
\eq{S_{JK} = \frac{1}{D}\ \sum_{p, q, k \in L_\Fus} d_k \bar{z}_{ppq}^{J;k} \bar{z}_{qqp}^{K;k}.
}
An anyon \(J\) has trivial self-statistics if \(\theta_J = 1\); two anyons \(J\) and \(K\) braid trivially if and only if \(S_{JK} = d_J d_K\), where \(d_J\) is the quantum dimension of anyon \(J\), defined as
\eq{
d_J = \sum_{J\text{'Charges } p} d_p.
}

States with two quasiparticles in two non-adjacent plaquettes are generated by ribbon operators along longer paths. These longer ribbon operators result from concatenating shorter ribbon operators. For example, to create two quasiparticles \(J^\ast\) and \(J\) with charges \(p^\ast_0\) and \(p_n\) in two non-adjacent plaquettes \(P_0\) and \(P_n\), we can choose a sequence of plaquettes \((P_0, P_1, \cdots, P_n)\), where \(P_i\) and \(P_{i+1}\) are adjacent plaquettes with their common edge \(E_i\). The ribbon operator \(W_{P_0P_n}^{J; p_0p_n}\) is
\eq{
W_{P_0P_n}^{J; p_0p_n} := \Bigg[\sum_{p_1 p_2 \cdots p_{n-1} \in L_\Fus} \prod_{k = 1}^{n-1} \left(d_{p_k} B_{P_k} W_{E_k}^{J; p_k p_{k+1}}\right)\Bigg]W_{E_0}^{J; p_0 p_1}.
}
Different choices of plaquette paths \((P_0, P_1, \cdots, P_n)\) give the same operator \(W_{P_0P_n}^{J; p_0 p_n}\) if these sequences can deform continuously from one to another. Following the same procedure, we can also define the creation operator of three or more anyons.

At the end of this section, we define the measurement operator \(M_P^J\) measuring whether there is an anyon \(J\) excited in plaquette \(P\):
\eqn{M_P^J\ \PlaquetteSrc
:= \sum_{s,t\in L_\Fus} \frac{d_sd_tz_{pps}^{J; t}}{d_p} \PlaquetteMsr\ .}
The set of measurement operators are orthonormal and complete:
\eq{M_P^JM_P^K = \delta_{JK}M_P^J,\qquad \sum_{J\in L_{\mathcal{Z}(\Fus)}}M_P^J = \idm.}

\subsection{The Output UMTC is the Center of the Input UFC}\label{sec:halfbraid}

As mentioned earlier, the string-net model's output UMTC \(\Cent(\Fus)\) is the center of its input UFC \(\Fus\), and the model is a physical representation of this center relationship. In this appendix, we explicitly demonstrate how this representation is understood.

Categorically, an object \(J\) in the center \(\Cent(\Fus)\) is denoted as a pair \(J = (X_J, c_{X_J, \cdot})\), where \(X_J\) is an object in UFC \(\Fus\), and \(c_{x_J, \cdot}\), called a \emph{half-braiding}, is a set of morphisms
\eq{
\{ c_{X_J, y}: X_J\otimes y\to y\otimes X_J | y\in \Fus
\}. }
A morphism \(c_{X_J, y}\) braids object \(X_J\) with object \(y\) in \(\Fus\) and can be depicted as
\eq{\HalfBraidingA.}
In a UFC, all morphisms can be decomposed as direct sums of fusion of simple objects, and so can a half-braiding:
\eqn[eq:halfB]{\HalfBraidingB.}
Here, \(L_J\subseteq L_\Fus\), such that the direct sum of simple objects in \(L_J\) is \(X_J\):
\eq{X_J = \bigoplus_{p \in L_J} p.}
The expansion coefficients \(z_{pqk}^{J;y}\) are known as the \emph{half-braiding tensor} of \(J\). A half-braiding should commute with any fusion in \(\Fus\):
\eqn[eq:halfC]{\HalfBraidingC.}
Expanding Eqs. \eqref{eq:halfC} using Eq. \eqref{eq:halfB} leads to Eq. \eqref{eq:halfA}.

For a string-net model with input UFC $\Fus$, an anyon type \(J\) is a simple object in the output UMTC \(\Cent(\Fus)\), and $J$'s charges take value in \(L_J\). The action of creation operator \(W_E^{J;pq}\) directly represents the half-braiding morphism \(c_{X_J, j_E}\) of object \(X_J\) with \(j_E\in L_\Fus\), the dof on edge \(E\):
\eq{\HalfBraidingD}

\subsection{The String-Net Model with Non-Commutative Input UFCs}\label{appendix:noab}

In this appendix, we briefly introduce the string-net model, whose input UFC $\Fus$ has non-commutative fusion rules $\delta$, i.e., there exists $a, b, c\in L_\Fus$, such that $\delta_{abc}\ne\delta_{acb}$. Note that the states in the Hilbert space of such a string-net model must satisfy the constraint \(\delta_{abc} \ne 0\) for any three incident edges or tails meeting at a vertex, \emph{counterclockwise} carrying degrees of freedom \(a, b, c\), and all pointing toward the vertex. 

For these non-commutative string-net models, Eq. \eqref{eq:halfA} for half-braiding tensors $z$ may admit matrix solutions. In other words, a minimal solution \( z_{pqj}^{J;k} \) of the half-braiding tensor may not be a complex number but instead a unitary matrix, where $J$ is the anyon types labeling distinct minimal solutions, and $p, q, j, k\in L_\Fus$. The $z^J$ tensor then becomes:
\eq{[z_{pqj}^{J;k}]_{\alpha\beta} \in \mathbb{C}, \qquad 1 \leq \alpha \leq n_{(J, p)},\qquad 1\le \beta \leq n_{(J, q)},}
where $n_{(J, p)}\in\mathbb{N}$. For those anyons $J$ with $n_{(J, p)} > 1$ for charge $p$, a dyon type is not only labeled by the anyon species $J$ but also by the multiplicity index $1 \le \alpha\le n_{(J,p)}$:
$$(J, p, \alpha),\qquad J\in L_{\Cent(\Fus)},\qquad p\text{ is }J\text{'s charge},\qquad 1\le\alpha\le n_{(J, p)}.$$
For our convenience, we refer to the pair \((J, p)\) of anyon type $J$ and charge type $p$ as a \emph{dyon multiplet}, which contains \(n_{(J, p)}\) distinct dyon types $(J, p, \alpha)$, where $\alpha$ is called the \emph{multiplet index}. The number \(n_{(J, p)}\) is called the \emph{degeneracy} of the dyon multiplet \((J, p)\). The pair \((p, \alpha)\), consisting of the charge type \(p\) and the multiplet index \(\alpha\), is referred to as the \emph{local dof} of the anyon \(J\).

In the Hilbert space of a non-commutative string-net model, two excited states, each with a dyon in the same plaquette having the same anyon types \( J \) and charge types \( p \) but different multiplet indices \( \alpha \), should be orthogonal excited states. Nevertheless, the original Hilbert space, whose local Hilbert space on each tail is spanned only by the charges $p\in L_\Fus$, is not capable of accommodating such a large number of distinct excited states. Therefore, to capture the full dyon spectrum, we have to \emph{enlarge} the local Hilbert space of each \emph{tail}, as in our previous work [ArXiv:2408.02664]. Dofs on edges do not need to be extended, as the edges pertain to the ground states because any path along an edge forms a closed loop, where each vertex only respects the fusion rules, disregarding the multiplet indices.

For convenience, the local subspace on each tail is spanned by local dofs of dyons, denoted as \( p_{J, \alpha} \), where \( J \) is an anyon species having charge \( p \), and \( 1 \leq \alpha \leq n_{(J,p)} \). Different basis states on a tail are orthogonal local states:
$$\Bigg\langle\quad\Tail{p_{J }}\ \Bigg|\quad\Tail{q_{K,\beta}}\ \Bigg\rangle = \delta_{JK}\delta_{pq}\delta_{\alpha\beta}.$$
Here, we enlarged the tails' local Hilbert spaces even for commutative string-net models. This does not affect our discussions, as two excited states with different anyon types in the same plaquettes must be orthogonal.

The simplest creation operator $W_E^{J;(p, \alpha)(q, \beta)}$ creating a pair of dyons $(J^\ast, p^\ast, \alpha)$ and $(J, q, \beta)$ in two adjacent plaquettes separated by edge $E$ is now defined as
\eqn{
W_E^{J; (p,\alpha)(q,\beta)} \NonAbA := \sum_{k \in L_\Fus} \sqrt{\frac{d_k}{d_j}}\ \ \overline{[{z_{pqj}^{J;k}}]_{\alpha\beta}} \ \ \NonAbB \ ,
}
Here, $\mathcal{I}\in L_{\Cent(\Fus)}$ is the trivial anyon with unique charge $1\in L_\Fus$ and has no degeneracy $n_\mathcal{I} = 1$. The creation operator multiplications depend on the CG coefficients of the tensor products of the half-braiding \( z \)-tensors. We will not detail them here.

A typical example of such a non-commutative input UFC is \({\tt Vec}(G)\), where \(G\) is a nonabelian group. The simple objects of UFC \({\tt Vec}(G)\) are labeled by group elements in $G$, and for any $a, b, c, d, e, f\in G$,
$$\delta_{abc} = 1\quad\text{if}\quad c^\ast = ab\quad\text{else}\quad 0,\qquad\qquad d_a = 1,\qquad\qquad G^{abe}_{cdf} = \delta_{abe}\delta_{bcf^\ast}\delta_{cde^\ast}\delta_{daf},$$
where $c^\ast$ is the inversed element of $c\in G$. Equation \eqref{eq:halfA} for $z$-tensors turns out to be
$$z^{J;(pa)}_{p(a^\ast pa)a}z^{J;(a^\ast pab)}_{(a^\ast pa)(b^\ast a^\ast pab)b} = z^{J;(pab)}_{p(b^\ast a^\ast pab)(ab)}.$$
Consequently, the anyon types $J$ are labeled by a pair
$$J = (\bar p, \rho),$$
where $\bar p = \{g^\ast pg| g\in G\}$ for $p\in G$ is a conjugate class of $G$ containing element $p$, and $\rho$ is an irreducible representation of the centralizer of conjugate class $\bar p$:
$$Z(\bar p) = \{g\in G| g^\ast pg = p\}.$$
The multiplet degeneracy of each dyon multiplet $(J, p)$ of anyon type $J$ is the dimension of irrep $\rho$ of centralizer $Z(\bar p)$:
$$n_{((\bar p, \rho), p)} = {\tt dim}\rho,$$
and the quantum dimension of anyon $J$ is $n_{((\bar p, \rho), p)}|\bar p|$, the number of all $J$'s local dofs.

In particular, the fluxons in the string-net model with input UFC ${\tt Vec}G$ are those dyons with anyon type $(\bar 1, \rho)$, whose trivial conjugate class is $\bar 1 = \{1\}\subset G$ and $Z(\bar 1) = G$. The correspoinding half-braiding tensor is:
$$z_{1,1,r}^{(\bar 1, \rho);r} = D_{\alpha\beta}^\rho(g),\qquad 1\le\alpha, \beta\le{\tt dim}\rho,\qquad\forall r\in G.$$
where $D^\rho$ is the representation matrix of irrep $\rho$ of group $G$.

\section{The Anyon Species of \eqs{S_3} Quantum Double Phase}\label{sec:anyon}

There are $8$ types of anyons in $D(S_3)$ phase, which is showed in Table \ref{tab:anyon species}.

Their corresponding $z$-tensors are:
\begin{equation}
    z^{A;1}_{1,1,1}=z^{A;r}_{1,1,r}=z^{A;r^2}_{1,1,r^2}=z^{A;s}_{1,1,s}=z^{A;rs}_{1,1,rs}=z^{A;sr}_{1,1,sr}=1;
\end{equation}
\begin{equation}
    z^{B;1}_{1,1,1}=z^{B;r}_{1,1,r}=z^{B;r^2}_{1,1,r^2}=1,\ z^{B;s}_{1,1,s}=z^{B;rs}_{1,1,rs}=z^{B;sr}_{1,1,sr}=-1;
\end{equation}
\begin{equation}
    \begin{split}
        &z^{C;1}_{1_1,1_1,1}=1,\ z^{C;r}_{1_1,1_1,r}=e^{i\frac{2\pi}{3}},\ z^{C;r^2}_{1_1,1_1,r^2}=e^{-i\frac{2\pi}{3}},\\
        &z^{C;1}_{1_2,1_2,1}=1,\ z^{C;r}_{1_2,1_2,r}=e^{-i\frac{2\pi}{3}},\ z^{C;r^2}_{1_2,1_2,r^2}=e^{i\frac{2\pi}{3}},\\
        &z^{C;s}_{1_1,1_2,s}=1,\ z^{C;rs}_{1_1,1_2,rs}=e^{i\frac{2\pi}{3}},\ z^{C;sr}_{1_1,1_2,sr}=e^{-i\frac{2\pi}{3}},\\
        &z^{C;s}_{1_2,1_1,s}=1,\ z^{C;rs}_{1_2,1_1,rs}=e^{-i\frac{2\pi}{3}},\ z^{C;sr}_{1_2,1_1,sr}=e^{i\frac{2\pi}{3}};\\
    \end{split}
\end{equation}
\begin{equation}
    \begin{split}
        &z^{D;s}_{s,s,1}=z^{D;sr}_{s,rs,r}=z^{D;rs}_{s,sr,r^2}=z^{D;rs}_{rs,rs,1}=z^{D;s}_{rs,sr,r}=z^{D;sr}_{rs,s,r^2}=1,\\
        &z^{D;sr}_{sr,sr,1}=z^{D;rs}_{sr,s,r}=z^{D;s}_{sr,rs,r^2}=z^{D;1}_{s,s,s}=z^{D;r}_{s,rs,sr}=z^{D;r^2}_{s,sr,rs}=1,\\
        &z^{D;1}_{rs,rs,rs}=z^{D;r}_{rs,sr,s}=z^{D;r^2}_{rs,s,sr}=z^{D;1}_{sr,sr,sr}=z^{D;r}_{sr,s,sr}=z^{D;r^2}_{sr,rs,s}=1;\\
    \end{split}
\end{equation}
\begin{equation}
    \begin{split}
        &z^{E;s}_{s,s,1}=z^{E;sr}_{s,rs,r}=z^{E;rs}_{s,sr,r^2}=z^{E;rs}_{rs,rs,1}=z^{E;s}_{rs,sr,r}=z^{E;sr}_{rs,s,r^2}=1,\\
        &z^{E;sr}_{sr,sr,1}=z^{E;rs}_{sr,s,r}=z^{E;s}_{sr,rs,r^2}=1,\ z^{E;1}_{s,s,s}=z^{E;r}_{s,rs,sr}=z^{E;r^2}_{s,sr,rs}=-1,\\
        &z^{E;1}_{rs,rs,rs}=z^{E;r}_{rs,sr,s}=z^{E;r^2}_{rs,s,sr}=z^{E;1}_{sr,sr,sr}=z^{E;r}_{sr,s,sr}=z^{E;r^2}_{sr,rs,s}=-1;\\
    \end{split}
\end{equation}
\begin{equation}
    \begin{split}
        &z^{F;r}_{r,r,1}=z^{F;r^2}_{r,r,r}=z^{F;1}_{r,r,r^2}=z^{F;r^2}_{r^2,r^2,1}=z^{F;1}_{r^2,r^2,r}=z^{F;r}_{r^2,r^2,r^2}=1,\\
        &z^{F;rs}_{r,r^2,s}=z^{F;sr}_{r,r^2,rs}=z^{F;rs}_{r,r^2,sr}=z^{F;sr}_{r^2,r,s}=z^{F;s}_{r^2,r,rs}=z^{F;rs}_{r^2,r,sr}=1;\\
    \end{split}
\end{equation}
\begin{equation}
    \begin{split}
        &z^{G;r}_{r,r,1}=1,\ z^{G;r^2}_{r,r,r}=e^{i\frac{2\pi}{3}},\ z^{G;1}_{r,r,r^2}=e^{-i\frac{2\pi}{3}},\\
        &z^{G;r^2}_{r^2,r^2,1}=1,\ z^{G;1}_{r^2,r^2,r}=e^{-i\frac{2\pi}{3}},\ z^{G;r}_{r^2,r^2,r^2}=e^{i\frac{2\pi}{3}},\\
        &z^{G;rs}_{r,r^2,s}=1,\ z^{G;sr}_{r,r^2,rs}=e^{i\frac{2\pi}{3}},\ z^{G;rs}_{r,r^2,sr}=e^{-i\frac{2\pi}{3}},\\
        &z^{G;sr}_{r^2,r,s}=1,\ z^{G;s}_{r^2,r,rs}=e^{-i\frac{2\pi}{3}},\ z^{G;rs}_{r^2,r,sr}=e^{i\frac{2\pi}{3}};\\
    \end{split}
\end{equation}
\begin{equation}
    \begin{split}
        &z^{H;r}_{r,r,1}=1,\ z^{H;r^2}_{r,r,r}=e^{-i\frac{2\pi}{3}},\ z^{H;1}_{r,r,r^2}=e^{i\frac{2\pi}{3}},\\
        &z^{H;r^2}_{r^2,r^2,1}=1,\ z^{H;1}_{r^2,r^2,r}=e^{i\frac{2\pi}{3}},\ z^{H;r}_{r^2,r^2,r^2}=e^{-i\frac{2\pi}{3}},\\
        &z^{H;rs}_{r,r^2,s}=1,\ z^{H;sr}_{r,r^2,rs}=e^{-i\frac{2\pi}{3}},\ z^{H;rs}_{r,r^2,sr}=e^{i\frac{2\pi}{3}},\\
        &z^{H;sr}_{r^2,r,s}=1,\ z^{H;s}_{r^2,r,rs}=e^{i\frac{2\pi}{3}},\ z^{H;rs}_{r^2,r,sr}=e^{-i\frac{2\pi}{3}}.\\
    \end{split}
\end{equation}

The $D(S_3)$ anyon braiding and self statistics are captured by the corresponding modular \(S\) and \(T\) matrices. From the $z$-tensors above, we can get the \(S\) and \(T\) matrices:
\begin{equation}
S=\frac{1}{6}\begin{pmatrix}
1 & 1 & 2 & 3 & 3 & 2 & 2 & 2\\
1 & 1 & 2 & -3 & -3 & 2 & 2 & 2\\
2 & 2 & 4 & 0 & 0 & -2 & -2 & -2\\
3 & -3 & 0 & 3 & -3 & 0 & 0 & 0\\
3 & -3 & 0 & -3 & 3 & 0 & 0 & 0\\
2 & 2 & -2 & 0 & 0 & 4 & -2 & -2\\
2 & 2 & -2 & 0 & 0 & -2 & -2 & 4\\
2 & 2 & -2 & 0 & 0 & -2 & 4 & -2\\
\end{pmatrix},\quad
T=\begin{pmatrix}
1 & 0 & 0 & 0 & 0 & 0 & 0 & 0\\
0 & 1 & 0 & 0 & 0 & 0 & 0 & 0\\
0 & 0 & 1 & 0 & 0 & 0 & 0 & 0\\
0 & 0 & 0 & 1 & 0 & 0 & 0 & 0\\
0 & 0 & 0 & 0 & -1 & 0 & 0 & 0\\
0 & 0 & 0 & 0 & 0 & 1 & 0 & 0\\
0 & 0 & 0 & 0 & 0 & 0 & e^{\frac{2\pi i}{3}} & 0\\
0 & 0 & 0 & 0 & 0 & 0 & 0 & e^{-\frac{2\pi i}{3}}\\
\end{pmatrix}.
\end{equation}
Note that any global symmetry permissible on a topological phase preserves the $S$ and $T$ matrices of the phase. Seen from the modular \(S\) and \(T\) matrices, the $D(S_3)$ phase admits a $Z_2$ symmetry that exchanges its anyon types $C$ and $F$, while preserving all other anyon types. Since the $C$-type anyon has only a trivial flux but a 2-dimensional charge space, while the $F$-type anyon has two flux type but trivial charge space, this $\Z_2$ symmetry is an EM-exchange symmetry.

\section{Multi-fusion Category as the Input of SET Model}\label{sec:multifusion}

In this section, we introduce how the multifusion category can be used as the input of SET string-net model.

\begin{figure}[t]
    \centering
    \FatLattice
    \caption{The fattened lattice of one plaquette.}
    \label{fig:fatlat}
\end{figure}

\subsection{Input data: A Multifusion Category with Associated Isomorphisms}

When a topological phase $\mathcal{C}$ is paired with a global symmetry $G$, it becomes an SET phase, denoted as $\mathcal{C}_G$. The SET phase $\mathcal{C}_G$ comprises multiple $G$-graded sectors, each of which is topologically equivalent to $\mathcal{C}$ but distinguished by a unique element from the symmetry group. The action of $G$ facilitates transitions between these $G$-graded sectors. This observation leads to the development of a lattice model for $\mathcal{C}_G$ by iterating the HGW string-net model of $\mathcal{C}$ a corresponding number of times. These iterations are interconnected through the elements of the group $G$, ensuring that in any basis configuration of the string-net model, each plaquette is associated with a specific $G$-graded sector. Given that the input data for the HGW model of phase $\mathcal{C}$ is a UFC $\Fus$, it follows that the input data for the lattice model associated with the $\mathcal{C}_G$ phase should similarly involve a proportional number of $\Fus$ copies, which are interconnected via the symmetry $G$. This configuration is mathematically formalized as a multifusion category $\M$ endowed with a $G$-action.

To grasp the concept of a multifusion category, consider a straightforward example where a multifusion category $\M$, derived from an $\Fus$, is generally expressed in matrix form:
\begin{equation}\label{eq:multifusionmatrix}
\M = \begin{pmatrix}
\Fus_{g_1g_1} & \Fus_{g_1g_2} & \cdots & \Fus_{g_1g_n} \\
\Fus_{g_2g_1} & \Fus_{g_2g_2} & \cdots & \Fus_{g_2g_n} \\
\vdots & \vdots & \ddots & \vdots \\
\Fus_{g_ng_1} & \Fus_{g_ng_2}& \cdots & \Fus_{g_ng_n}
\end{pmatrix},
\end{equation}
Each $\Fus_{g_ig_j}$ can be considered as an instance of $\Fus$ when viewed as a UFC. Generally, the size $n$ corresponds to $|G|$, considering $G$ is finite. A simple object within $\M$ is symbolized by $x_{g_ig_j}$, which also represents a simple object within $\Fus_{g_ig_j}$, where $x$ is a simple object within \(\Fus\). These simple objects adhere to the subsequent fusion rules:
\eqn[eq:mulfusrule]{
x_{g_ig_j}\otimes y_{g_kg_l} = \begin{cases}(x\otimes y)_{g_ig_l}\quad\ (g_j =g_ k),\\ \nulm\qquad\qquad\quad(g_j\ne g_k)\end{cases},
}
where $\nulm$ is the zero object of \(\M\), and $x\otimes y$ is a fusion in $\Fus$. 

According to the fusion rules, the lattice illustrated in Figure \ref{fig:lat} can be expanded by transforming each edge or tail into a double line, whereby the indices of the group elements reside on the double lines \cite{Chang2015}, as shown in Figure \ref{fig:fatlat}. It is evident that each plaquette can now be uniquely identified by a group element $g$, with $g$ serving as the label for the loop of the inner-side line within the respective plaquette. We refer to the plaquette as being in a $g$-sector. This method is equivalent to introducing spins into the plaquettes \cite{heinrich2016symmetry}.

The simple objects corresponding to the diagonal elements $\Fus_{gg}$ of the multifusion matrix \eqref{eq:multifusionmatrix} characterize the degrees of freedom (dofs) in sector-$g$. In contrast, the simple objects associated with the off-diagonal elements characterize the domain wall degrees of freedom. Specifically, the simple objects within $\Fus_{g_ig_j}$ describe the domain wall between sector-$g_i$ and sector-$g_j$.

Considering the structure of the input data, the lattice configuration of our SET model for phase $\mathcal{C}_G$ mirrors precisely that of the HGW string-net model for phase $\mathcal{C}$. This lattice is a honeycomb arrangement, where each plaquette is represented with a tail (a wiggly dangling edge), as depicted in Figure \ref{fig:lat}. Each edge and tail of the lattice embodies a degree of freedom that assumes values from the simple objects of $\M$, thereby ensuring that the fusion rules of $M$ are satisfied at every vertex within the lattice.

\subsection{Hamiltonian}

The Hamiltonian takes the form
\eqn[eq:hamiltonian]{
H=-\sum_{p}Q_p=-\sum_{p}\sum_{s}Q_p^{s},
}
where $p$ represents a summation over all plaquettes and $s$ encompasses all simple objects in the multifusion category $\mathcal{M}$. The Hamiltonian here is identical to the one described in Ref. \cite{zhao2024b}, which provides a reformulation of the Hamiltonian for the HGW string-net model, ensuring uniform energy levels for all elementary anyon excitations. The operators $Q_p^{s}$ are discussed in detail in Appendix \ref{sec:review}.

While Hamiltonian \eqref{eq:hamiltonian} theoretically aligns with the HGW string-net model representing a pure topological phase, the physical phenomena it represents diverges notably from that of the HGW model. Specifically, the system incorporates both anyon excitations, known as anyons, and local excitations, which the HGW model lacks. In the Symmetry-Enriched Topological (SET) phase characterized by Hamiltonian \eqref{eq:hamiltonian}, the anyon excitations are defined in respect to the local excitations, which should be considered as part of the vacuum. This phenomenon, referred to as symmetry fractionalization, arises due to the nontrivial action of global symmetry on the topological vacuum, which includes local excitations. Consequently, this interaction results in distinct sectors graded by the symmetry group $G$. Through the fusion rules of anyons, specific types of anyons are permitted to reside within a $G$-graded sector where they carry fractionalized symmetry charges. It is important to note that the ground-state degeneracy of a symmetry-enriched topological (SET) phase, denoted as $\mathcal{C}_G$, remains equivalent to that of the purely topological phase $\mathcal{C}$ for any given topology, notwithstanding the presence of $G$-graded sectors.

\section{The Categorical Data of \eqs{\Z_2} Symmetry Enriched \eqs{S_3} Multi-fusion Category}\label{sec:categorical_data}

The input multifusion category of $Z_2$ symmetry enriched $S_3$ string-net model can be represented by the multifusion matrix
\begin{equation}\label{eq:s3multifusion}
    \mathcal{M}=\left(\begin{array}{cc}
        \{1_{++},r_{++},r^2_{++},s_{++},rs_{++},sr_{++}\} & \{\alpha_{+-},\ \beta_{+-}\} \\
        \{\alpha_{-+},\ \beta_{-+}\} & \{1_{--},r_{--},r^2_{--},s_{--},rs_{--},sr_{--}\}
    \end{array}\right).
\end{equation}

In this section, we will list the categorical data of this multifusion category.

\subsection{The Fusion Rules}

The fusion rules are
\eqn{
&\delta_{M_KM_LM_N}=\delta_{kln},\ \forall M_K,M_L,M_N\in\{1_{--},r_{++},r^2_{++},s_{++},rs_{++},sr_{++}\},\\
&\delta_{\alpha_{-+}1_{++}\alpha_{+-}}=\delta_{\alpha_{-+}r_{++}\alpha_{+-}}=\delta_{\alpha_{-+}r^2_{++}\alpha_{+-}}=\delta_{\beta_{-+}1_{++}\beta_{+-}}=\delta_{\beta_{-+}r_{++}\beta_{+-}}=\delta_{\beta_{-+}r^2_{++}\beta_{+-}}=1,\\
&\delta_{\alpha_{-+}s_{++}\beta_{+-}}=\delta_{\alpha_{-+}rs_{++}\beta_{+-}}=\delta_{\alpha_{-+}sr_{++}\beta_{+-}}=\delta_{\beta_{-+}s_{++}\alpha_{+-}}=\delta_{\beta_{-+}rs_{++}\alpha_{+-}}=\delta_{\beta_{-+}sr_{++}\alpha_{+-}}=1,\\
&\delta_{\alpha_{-+}\alpha_{+-}1_{--}}=\delta_{\alpha_{-+}\alpha_{+-}r_{--}}=\delta_{\alpha_{-+}\alpha_{+-}r^2_{--}}=\delta_{\beta_{-+}\beta_{+-}1_{--}}=\delta_{\beta_{-+}\beta_{+-}r_{--}}=\delta_{\beta_{-+}\beta_{+-}r^2_{--}}=1,\\
&\delta_{\alpha_{-+}\beta_{+-}s_{--}}=\delta_{\alpha_{-+}\beta_{+-}rs_{--}}=\delta_{\alpha_{-+}\beta_{+-}sr_{--}}=\delta_{\beta_{-+}\alpha_{+-}s_{--}}=\delta_{\beta_{-+}\alpha_{+-}rs_{--}}=\delta_{\beta_{-+}\alpha_{+-}sr_{--}}=1.\\
}

\subsection{The \eqs{6j}-Symbols}

The non-zero $6j$-symbols are
\begin{equation*}
    G^{\alpha_{+-}\alpha_{-+}1_{++}}_{\alpha_{+-}\alpha_{-+}1_{--}}=G^{\alpha_{+-}\alpha_{-+}1_{++}}_{\alpha_{+-}\alpha_{-+}r_{--}}=G^{\alpha_{+-}\alpha_{-+}1_{++}}_{\alpha_{+-}\alpha_{-+}r^2_{--}}=\frac{1}{\sqrt{3}},
\end{equation*}
\begin{equation*}
    G^{\alpha_{+-}\alpha_{-+}r_{++}}_{\alpha_{+-}\alpha_{-+}1_{--}}=\frac{1}{\sqrt{3}},\ G^{\alpha_{+-}\alpha_{-+}r_{++}}_{\alpha_{+-}\alpha_{-+}r_{--}}=\frac{e^{-i\frac{2\pi}{3}}}{\sqrt{3}},\ G^{\alpha_{+-}\alpha_{-+}r_{++}}_{\alpha_{+-}\alpha_{-+}r^2_{--}}=\frac{e^{i\frac{2\pi}{3}}}{\sqrt{3}},
\end{equation*}
\begin{equation*}
    G^{\alpha_{+-}\alpha_{-+}r^2_{++}}_{\alpha_{+-}\alpha_{-+}1_{--}}=\frac{1}{\sqrt{3}},\ G^{\alpha_{+-}\alpha_{-+}r^2_{++}}_{\alpha_{+-}\alpha_{-+}r_{--}}=\frac{e^{i\frac{2\pi}{3}}}{\sqrt{3}},\ G^{\alpha_{+-}\alpha_{-+}r^2_{++}}_{\alpha_{+-}\alpha_{-+}r^2_{--}}=\frac{e^{-i\frac{2\pi}{3}}}{\sqrt{3}},
\end{equation*}
\begin{equation*}
    G^{\beta_{+-}\beta_{-+}1_{++}}_{\beta_{+-}\beta_{-+}1_{--}}=G^{\beta_{+-}\beta_{-+}1_{++}}_{\beta_{+-}\beta_{-+}r_{--}}=G^{\beta_{+-}\beta_{-+}1_{++}}_{\beta_{+-}\beta_{-+}r^2_{--}}=\frac{1}{\sqrt{3}},
\end{equation*}
\begin{equation*}
    G^{\beta_{+-}\beta_{-+}r_{++}}_{\beta_{+-}\beta_{-+}1_{--}}=\frac{1}{\sqrt{3}},\ G^{\beta_{+-}\beta_{-+}r_{++}}_{\beta_{+-}\beta_{-+}r_{--}}=\frac{e^{i\frac{2\pi}{3}}}{\sqrt{3}},\ G^{\beta_{+-}\beta_{-+}r_{++}}_{\beta_{+-}\beta_{-+}r^2_{--}}=\frac{e^{-i\frac{2\pi}{3}}}{\sqrt{3}},
\end{equation*}
\begin{equation*}
    G^{\beta_{+-}\beta_{-+}r^2_{++}}_{\beta_{+-}\beta_{-+}1_{--}}=\frac{1}{\sqrt{3}},\ G^{\beta_{+-}\beta_{-+}r^2_{++}}_{\beta_{+-}\beta_{-+}r_{--}}=\frac{e^{-i\frac{2\pi}{3}}}{\sqrt{3}},\ G^{\beta_{+-}\beta_{-+}r^2_{++}}_{\beta_{+-}\beta_{-+}r^2_{--}}=\frac{e^{i\frac{2\pi}{3}}}{\sqrt{3}},
\end{equation*}
\begin{equation*}
    G^{\alpha_{+-}\beta_{-+}s_{++}}_{\beta_{+-}\alpha_{-+}1_{--}}=G^{\alpha_{+-}\beta_{-+}s_{++}}_{\beta_{+-}\alpha_{-+}r_{--}}=G^{\alpha_{+-}\beta_{-+}s_{++}}_{\beta_{+-}\alpha_{-+}r^2_{--}}=\frac{1}{\sqrt{3}},
\end{equation*}
\begin{equation*}
    G^{\alpha_{+-}\beta_{-+}rs_{++}}_{\beta_{+-}\alpha_{-+}1_{--}}=\frac{1}{\sqrt{3}},\ G^{\alpha_{+-}\beta_{-+}rs_{++}}_{\beta_{+-}\alpha_{-+}r_{--}}=\frac{e^{i\frac{2\pi}{3}}}{\sqrt{3}},\ G^{\alpha_{+-}\beta_{-+}rs_{++}}_{\beta_{+-}\alpha_{-+}r^2_{--}}=\frac{e^{-i\frac{2\pi}{3}}}{\sqrt{3}},
\end{equation*}
\begin{equation*}
    G^{\alpha_{+-}\beta_{-+}sr_{++}}_{\beta_{+-}\alpha_{-+}1_{--}}=\frac{1}{\sqrt{3}},\ G^{\alpha_{+-}\beta_{-+}sr_{++}}_{\beta_{+-}\alpha_{-+}r_{--}}=\frac{e^{-i\frac{2\pi}{3}}}{\sqrt{3}},\ G^{\alpha_{+-}\beta_{-+}sr_{++}}_{\beta_{+-}\alpha_{-+}r^2_{--}}=\frac{e^{i\frac{2\pi}{3}}}{\sqrt{3}},
\end{equation*}
\begin{equation*}
    G^{\alpha_{+-}\alpha_{-+}1_{++}}_{\beta_{+-}\beta_{-+}s_{--}}=G^{\alpha_{+-}\alpha_{-+}1_{++}}_{\beta_{+-}\beta_{-+}rs_{--}}=G^{\alpha_{+-}\alpha_{-+}1_{++}}_{\beta_{+-}\beta_{-+}sr_{--}}=\frac{1}{\sqrt{3}},
\end{equation*}
\begin{equation*}
    G^{\alpha_{+-}\alpha_{-+}r_{++}}_{\beta_{+-}\beta_{-+}s_{--}}=\frac{1}{\sqrt{3}},\ G^{\alpha_{+-}\alpha_{-+}r_{++}}_{\beta_{+-}\beta_{-+}rs_{--}}=\frac{e^{-i\frac{2\pi}{3}}}{\sqrt{3}},\ G^{\alpha_{+-}\alpha_{-+}r_{++}}_{\beta_{+-}\beta_{-+}sr_{--}}=\frac{e^{i\frac{2\pi}{3}}}{\sqrt{3}},
\end{equation*}
\begin{equation*}
    G^{\alpha_{+-}\alpha_{-+}r^2_{++}}_{\beta_{+-}\beta_{-+}s_{--}}=\frac{1}{\sqrt{3}},\ G^{\alpha_{+-}\alpha_{-+}r^2_{++}}_{\beta_{+-}\beta_{-+}rs_{--}}=\frac{e^{i\frac{2\pi}{3}}}{\sqrt{3}},\ G^{\alpha_{+-}\alpha_{-+}r^2_{++}}_{\beta_{+-}\beta_{-+}sr_{--}}=\frac{e^{-i\frac{2\pi}{3}}}{\sqrt{3}},
\end{equation*}
\begin{equation*}
    G^{\alpha_{+-}\beta_{-+}s_{++}}_{\alpha_{+-}\beta_{-+}s_{--}}=G^{\alpha_{+-}\beta_{-+}s_{++}}_{\alpha_{+-}\beta_{-+}rs_{--}}=G^{\alpha_{+-}\beta_{-+}s_{++}}_{\alpha_{+-}\beta_{-+}sr_{--}}=\frac{1}{\sqrt{3}},
\end{equation*}
\begin{equation*}
    G^{\alpha_{+-}\beta_{-+}rs_{++}}_{\alpha_{+-}\beta_{-+}s_{--}}=\frac{1}{\sqrt{3}},\ G^{\alpha_{+-}\beta_{-+}rs_{++}}_{\alpha_{+-}\beta_{-+}rs_{--}}=\frac{e^{i\frac{2\pi}{3}}}{\sqrt{3}},\ G^{\alpha_{+-}\beta_{-+}rs_{++}}_{\alpha_{+-}\beta_{-+}sr_{--}}=\frac{e^{-i\frac{2\pi}{3}}}{\sqrt{3}},
\end{equation*}
\begin{equation*}
    G^{\alpha_{+-}\beta_{-+}sr_{++}}_{\alpha_{+-}\beta_{-+}s_{--}}=\frac{1}{\sqrt{3}},\ G^{\alpha_{+-}\beta_{-+}sr_{++}}_{\alpha_{+-}\beta_{-+}rs_{--}}=\frac{e^{-i\frac{2\pi}{3}}}{\sqrt{3}},\ G^{\alpha_{+-}\beta_{-+}sr_{++}}_{\alpha_{+-}\beta_{-+}sr_{--}}=\frac{e^{i\frac{2\pi}{3}}}{\sqrt{3}},
\end{equation*}
\begin{equation*}
    G^{abc}_{\mu\nu^*\gamma^*}=\frac{1}{\sqrt[4]{3}}\delta_{abc}\delta_{\nu^*c\mu}\delta_{\gamma^*a\nu}\delta_{\mu^*b\gamma},
\end{equation*}
\begin{equation*}
    \forall a,b,c\in\{1_{++},r_{++},r^2_{++},s_{++},rs_{++},sr_{++}\},\ \mu,\nu,\gamma\in \{\alpha_{+-},\beta_{+-}\},
\end{equation*}
\begin{equation*}
    G^{abc}_{\mu^*\nu\gamma}=\frac{1}{\sqrt[4]{3}}\delta_{abc}\delta_{\nu c\mu^*}\delta_{\gamma a\nu^*}\delta_{\mu b\gamma^*},
\end{equation*}
\begin{equation*}
    \forall a,b,c\in\{1_{--},r_{--},r^2_{--},s_{--},rs_{--},sr_{--}\},\ \mu,\nu,\gamma\in \{\alpha_{+-},\beta_{+-}\}.
\end{equation*}

\section{The anyon Spectrum of EM-Exchange Symmetry Enriched \eqs{S_3} String-net Model}\label{sec:symmetry_anyon}

In this section, we provide the topological excitaion spectrum of EM-exchange symmetry enriched $S_3$ string-net model. There are $8$ kinds of anyon excitations, each one can be denoted by a pair $(X,Y)$, which means that it appears as the original $D(S_3)$ $X$-type ($Y$-type) anyon in the $+$ ($-$) sector. The $z$-tensors of these excitations are:
\begin{align*}
        &z^{(A,A);1_{++}}_{1_{++},1_{++},1_{++}}=z^{(A,A);r_{++}}_{1_{++},1_{++},r_{++}}=z^{(A,A);r^2_{++}}_{1_{++},1_{++},r^2_{++}}=z^{(A,A);s_{++}}_{1_{++},1_{++},s_{++}}=1,\\
        &z^{(A,A);rs_{++}}_{1_{++},1_{++},rs_{++}}=z^{(A,A);sr_{++}}_{1_{++},1_{++},sr_{++}}=z^{(A,A);\alpha_{+-}}_{1_{++},1_{--},\alpha_{+-}}=z^{(A,A);\beta_{+-}}_{1_{++},1_{--},\beta_{+-}}=1,\\
        &z^{(A,A);1_{--}}_{1_{--},1_{--},1_{--}}=z^{(A,A);r_{--}}_{1_{--},1_{--},r_{--}}=z^{(A,A);r^2_{--}}_{1_{--},1_{--},r^2_{--}}=z^{(A,A);s_{--}}_{1_{--},1_{--},s_{--}}=1,\\
        &z^{(A,A);rs_{--}}_{1_{--},1_{--},rs_{--}}=z^{(A,A);sr_{--}}_{1_{--},1_{--},sr_{--}}=z^{(A,A);\alpha_{-+}}_{1_{--},1_{++},\alpha_{-+}}=z^{(A,A);\beta_{-+}}_{1_{--},1_{++},\beta_{-+}}=1;
\end{align*}
\begin{align*}
        &z^{(B,B);1_{++}}_{1_{++},1_{++},1_{++}}=z^{(B,B);r_{++}}_{1_{++},1_{++},r_{++}}=z^{(B,B);r^2_{++}}_{1_{++},1_{++},r^2_{++}}=1,\\
        &z^{(B,B);s_{++}}_{1_{++},1_{++},s_{++}}=z^{(B,B);rs_{++}}_{1_{++},1_{++},rs_{++}}=z^{(B,B);sr_{++}}_{1_{++},1_{++},sr_{++}}=-1,\\
        &z^{(B,B);1_{--}}_{1_{--},1_{--},1_{--}}=z^{(B,B);r_{++}}_{1_{--},1_{--},r_{++}}=z^{(B,B);r^2_{++}}_{1_{--},1_{--},r^2_{++}}=1,\\
        &z^{(B,B);s_{--}}_{1_{--},1_{--},s_{--}}=z^{(B,B);rs_{--}}_{1_{--},1_{--},rs_{--}}=z^{(B,B);sr_{--}}_{1_{--},1_{--},sr_{--}}=-1,\\
        &z^{(B,B);\alpha_{+-}}_{1_{--},1_{--},\alpha_{+-}}=z^{(B,B);\beta_{+-}}_{1_{--},1_{--},\beta_{+-}}=z^{(B,B);\alpha_{-+}}_{1_{--},1_{++},\alpha_{-+}}=z^{(B,B);\beta_{-+}}_{1_{--},1_{++},\beta_{-+}}=1;
\end{align*}
\begin{align*}
        &z^{(C,F);1_{++}}_{(1_1)_{++},(1_1)_{++},1_{++}}=1,\ z^{(C,F);r_{++}}_{(1_1)_{++},(1_1)_{++},r_{++}}=e^{i\frac{2\pi}{3}},\ z^{(C,F);r^2_{++}}_{(1_1)_{++},(1_1)_{++},r^2_{++}}=e^{-i\frac{2\pi}{3}},\\
        &z^{(C,F);1_{++}}_{(1_2)_{++},(1_2)_{++},1_{++}}=1,\ z^{(C,F);r_{++}}_{(1_2)_{++},(1_2)_{++},r_{++}}=e^{-i\frac{2\pi}{3}},\ z^{(C,F);r^2_{++}}_{(1_2)_{++},(1_2)_{++},r^2_{++}}=e^{i\frac{2\pi}{3}},\\
        &z^{(C,F);s_{++}}_{(1_1)_{++},(1_2)_{++},s_{++}}=1,\ z^{(C,F);rs_{++}}_{(1_1)_{++},(1_2)_{++},rs_{++}}=e^{i\frac{2\pi}{3}},\ z^{(C,F);sr_{++}}_{(1_1)_{++},(1_2)_{++},sr_{++}}=e^{-i\frac{2\pi}{3}},\\
        &z^{(C,F);s_{++}}_{(1_2)_{++},(1_1)_{++},s_{++}}=1,\ z^{(C,F);rs_{++}}_{(1_2)_{++},(1_1)_{++},rs_{++}}=e^{-i\frac{2\pi}{3}},\ z^{(C,F);sr_{++}}_{(1_2)_{++},(1_1)_{++},sr_{++}}=e^{i\frac{2\pi}{3}},\\
        &z^{(C,F);r_{--}}_{r_{--},r_{--},1_{--}}=z^{(C,F);r^2_{--}}_{r_{--},r_{--},r_{--}}=z^{(C,F);1_{--}}_{r_{--},r_{--},r^2_{--}}=1,\\
        &z^{(C,F);r^2_{--}}_{r^2_{--},r^2_{--},1_{--}}=z^{(C,F);1_{--}}_{r^2_{--},r^2_{--},r_{--}}=z^{(C,F);r_{--}}_{r^2_{--},r^2_{--},r^2_{--}}=1,\\
        &z^{(C,F);rs_{--}}_{r_{--},r^2_{--},s_{--}}=z^{(C,F);sr_{--}}_{r_{--},r^2_{--},rs_{--}}=z^{(C,F);rs_{--}}_{r_{--},r^2_{--},sr_{--}}=1,\\
        &z^{(C,F);sr_{--}}_{r^2_{--},r_{--},s_{--}}=z^{(C,F);s_{--}}_{r^2_{--},r_{--},rs_{--}}=z^{(C,F);rs_{--}}_{r^2_{--},r_{--},sr_{--}}=1,\\
        &z^{(C,F);\alpha_{+-}}_{(1_1)_{++},r_{--},\alpha_{+-}}=z^{(C,F);\alpha_{+-}}_{(1_2)_{++},r^2_{--},\alpha_{+-}}=z^{(C,F);\alpha_{-+}}_{r_{--},(1_1)_{++},\alpha_{-+}}=z^{(C,F);\alpha_{-+}}_{r^2_{--},(1_2)_{++},\alpha_{-+}}=1,\\
        &z^{(C,F);\beta_{+-}}_{(1_1)_{++},r^2_{--},\beta_{+-}}=z^{(C,F);\beta_{+-}}_{(1_2)_{++},r_{--},\beta_{+-}}=z^{(C,F);\beta_{-+}}_{r^2_{--},(1_1)_{++},\beta_{-+}}=z^{(C,F);\beta_{-+}}_{r_{--},(1_2)_{++},\beta_{-+}}=1;
\end{align*}
\begin{align*}
        &z^{(F,C);1_{--}}_{(1_1)_{--},(1_1)_{--},1_{--}}=1,\ z^{(F,C);r_{--}}_{(1_1)_{--},(1_1)_{--},r_{--}}=e^{i\frac{2\pi}{3}},\ z^{(F,C);r^2_{--}}_{(1_1)_{--},(1_1)_{--},r^2_{--}}=e^{-i\frac{2\pi}{3}},\\
        &z^{(F,C);1_{--}}_{(1_2)_{--},(1_2)_{--},1_{--}}=1,\ z^{(F,C);r_{--}}_{(1_2)_{--},(1_2)_{--},r_{--}}=e^{-i\frac{2\pi}{3}},\ z^{(F,C);r^2_{--}}_{(1_2)_{--},(1_2)_{--},r^2_{--}}=e^{i\frac{2\pi}{3}},\\
        &z^{(F,C);s_{--}}_{(1_1)_{--},(1_2)_{--},s_{--}}=1,\ z^{(F,C);rs_{--}}_{(1_1)_{--},(1_2)_{--},rs_{--}}=e^{i\frac{2\pi}{3}},\ z^{(F,C);sr_{--}}_{(1_1)_{--},(1_2)_{--},sr_{--}}=e^{-i\frac{2\pi}{3}},\\
        &z^{(F,C);s_{--}}_{(1_2)_{--},(1_1)_{--},s_{--}}=1,\ z^{(F,C);rs_{--}}_{(1_2)_{--},(1_1)_{--},rs_{--}}=e^{-i\frac{2\pi}{3}},\ z^{(F,C);sr_{--}}_{(1_2)_{--},(1_1)_{--},sr_{--}}=e^{i\frac{2\pi}{3}},\\
        &z^{(F,C);r_{++}}_{r_{++},r_{++},1_{++}}=z^{(F,C);r^2_{++}}_{r_{++},r_{++},r_{++}}=z^{(F,C);1_{++}}_{r_{++},r_{++},r^2_{++}}=1,\\
        &z^{(F,C);r^2_{++}}_{r^2_{++},r^2_{++},1_{++}}=z^{(F,C);1_{++}}_{r^2_{++},r^2_{++},r_{++}}=z^{(F,C);r_{++}}_{r^2_{++},r^2_{++},r^2_{++}}=1,\\
        &z^{(F,C);rs_{++}}_{r_{++},r^2_{++},s_{++}}=z^{(F,C);sr_{++}}_{r_{++},r^2_{++},rs_{++}}=z^{(F,C);rs_{++}}_{r_{++},r^2_{++},sr_{++}}=1,\\
        &z^{(F,C);sr_{++}}_{r^2_{++},r_{++},s_{++}}=z^{(F,C);s_{++}}_{r^2_{++},r_{++},rs_{++}}=z^{(F,C);rs_{++}}_{r^2_{++},r_{++},sr_{++}}=1,\\
        &z^{(F,C);\alpha_{-+}}_{(1_1)_{--},r_{++},\alpha_{-+}}=z^{(F,C);\alpha_{-+}}_{(1_2)_{--},r^2_{++},\alpha_{-+}}=z^{(F,C);\alpha_{+-}}_{r_{++},(1_1)_{--},\alpha_{+-}}=z^{(F,C);\alpha_{+-}}_{r^2_{++},(1_2)_{--},\alpha_{+-}}=1,\\
        &z^{(F,C);\beta_{-+}}_{(1_1)_{--},r^2_{++},\beta_{-+}}=z^{(F,C);\beta_{-+}}_{(1_2)_{--},r_{++},\beta_{-+}}=z^{(F,C);\beta_{+-}}_{r^2_{++},(1_1)_{--},\beta_{+-}}=z^{(F,C);\beta_{+-}}_{r_{++},(1_2)_{--},\beta_{+-}}=1;
\end{align*}
\begin{align*}
        &z^{(G,G);r_{++}}_{r_{++},r_{++},1_{++}}=1,\ z^{(G,G);r^2_{++}}_{r_{++},r_{++},r_{++}}=e^{i\frac{2\pi}{3}},\ z^{(G,G);1_{++}}_{r_{++},r_{++},r^2_{++}}=e^{-i\frac{2\pi}{3}},\\
        &z^{(G,G);r^2_{++}}_{r^2_{++},r^2_{++},1_{++}}=1,\ z^{(G,G);1_{++}}_{r^2_{++},r^2_{++},r_{++}}=e^{-i\frac{2\pi}{3}},\ z^{(G,G);r_{++}}_{r^2_{++},r^2_{++},r^2_{++}}=e^{i\frac{2\pi}{3}},\\
        &z^{(G,G);rs_{++}}_{r_{++},r^2_{++},s_{++}}=1,\ z^{(G,G);sr_{++}}_{r_{++},r^2_{++},rs_{++}}=e^{i\frac{2\pi}{3}},\ z^{(G,G);rs_{++}}_{r_{++},r^2_{++},sr_{++}}=e^{-i\frac{2\pi}{3}},\\
        &z^{(G,G);sr_{++}}_{r^2_{++},r_{++},s_{++}}=1,\ z^{(G,G);s_{++}}_{r^2_{++},r_{++},rs_{++}}=e^{-i\frac{2\pi}{3}},\ z^{(G,G);rs_{++}}_{r^2_{++},r_{++},sr_{++}}=e^{i\frac{2\pi}{3}},\\
        &z^{(G,G);r_{--}}_{r_{--},r_{--},1_{--}}=1,\ z^{(G,G);r^2_{--}}_{r_{--},r_{--},r_{--}}=e^{i\frac{2\pi}{3}},\ z^{(G,G);1_{--}}_{r_{--},r_{--},r^2_{--}}=e^{-i\frac{2\pi}{3}},\\
        &z^{(G,G);r^2_{--}}_{r^2_{--},r^2_{--},1_{--}}=1,\ z^{(G,G);1_{--}}_{r^2_{--},r^2_{--},r_{--}}=e^{-i\frac{2\pi}{3}},\ z^{(G,G);r_{--}}_{r^2_{--},r^2_{--},r^2_{--}}=e^{i\frac{2\pi}{3}},\\
        &z^{(G,G);rs_{--}}_{r_{--},r^2_{--},s_{--}}=1,\ z^{(G,G);sr_{--}}_{r_{--},r^2_{--},rs_{--}}=e^{i\frac{2\pi}{3}},\ z^{(G,G);rs_{--}}_{r_{--},r^2_{--},sr_{--}}=e^{-i\frac{2\pi}{3}},\\
        &z^{(G,G);sr_{--}}_{r^2_{--},r_{--},s_{--}}=1,\ z^{(G,G);s_{--}}_{r^2_{--},r_{--},rs_{--}}=e^{-i\frac{2\pi}{3}},\ z^{(G,G);rs_{--}}_{r^2_{--},r_{--},sr_{--}}=e^{i\frac{2\pi}{3}}, \\
        &z^{(G,G);\alpha_{+-}}_{r_{++},r_{--},\alpha_{+-}}=z^{(G,G);\alpha_{+-}}_{r^2_{++},r^2_{--},\alpha_{+-}}=z^{(G,G);\beta_{+-}}_{r_{++},r^2_{--},\beta_{+-}}=z^{(G,G);\beta_{+-}}_{r^2_{++},r_{--},\beta_{+-}}=e^{i\frac{2\pi}{3}},\\
        &z^{(G,G);\alpha_{-+}}_{r_{--},r_{++},\alpha_{-+}}=z^{(G,G);\alpha_{-+}}_{r^2_{--},r^2_{++},\alpha_{-+}}=z^{(G,G);\beta_{-+}}_{r_{--},r^2_{++},\beta_{-+}}=z^{(G,G);\beta_{-+}}_{r^2_{--},r_{++},\beta_{-+}}=e^{i\frac{2\pi}{3}};
\end{align*}

\begin{align*}
       &z^{(H,H);r_{++}}_{r_{++},r_{++},1_{++}}=1,\ z^{(H,H);r^2_{++}}_{r_{++},r_{++},r_{++}}=e^{-i\frac{2\pi}{3}},\ z^{(H,H);1_{++}}_{r_{++},r_{++},r^2_{++}}=e^{i\frac{2\pi}{3}},\\
        &z^{(H,H);r^2_{++}}_{r^2_{++},r^2_{++},1_{++}}=1,\ z^{(H,H);1_{++}}_{r^2_{++},r^2_{++},r_{++}}=e^{i\frac{2\pi}{3}},\ z^{(H,H);r_{++}}_{r^2_{++},r^2_{++},r^2_{++}}=e^{-i\frac{2\pi}{3}},\\
        &z^{(H,H);rs_{++}}_{r_{++},r^2_{++},s_{++}}=1,\ z^{(H,H);sr_{++}}_{r_{++},r^2_{++},rs_{++}}=e^{-i\frac{2\pi}{3}},\ z^{(H,H);rs_{++}}_{r_{++},r^2_{++},sr_{++}}=e^{i\frac{2\pi}{3}},\\
        &z^{(H,H);sr_{++}}_{r^2_{++},r_{++},s_{++}}=1,\ z^{(H,H);s_{++}}_{r^2_{++},r_{++},rs_{++}}=e^{i\frac{2\pi}{3}},\ z^{(H,H);rs_{++}}_{r^2_{++},r_{++},sr_{++}}=e^{-i\frac{2\pi}{3}},\\
        &z^{(H,H);r_{--}}_{r_{--},r_{--},1_{--}}=1,\ z^{(H,H);r^2_{--}}_{r_{--},r_{--},r_{--}}=e^{-i\frac{2\pi}{3}},\ z^{(H,H);1_{--}}_{r_{--},r_{--},r^2_{--}}=e^{i\frac{2\pi}{3}},\\
        &z^{(H,H);r^2_{--}}_{r^2_{--},r^2_{--},1_{--}}=1,\ z^{(H,H);1_{--}}_{r^2_{--},r^2_{--},r_{--}}=e^{i\frac{2\pi}{3}},\ z^{(H,H);r_{--}}_{r^2_{--},r^2_{--},r^2_{--}}=e^{-i\frac{2\pi}{3}}, \\
        &z^{(H,H);rs_{--}}_{r_{--},r^2_{--},s_{--}}=1,\ z^{(H,H);sr_{--}}_{r_{--},r^2_{--},rs_{--}}=e^{-i\frac{2\pi}{3}},\ z^{(H,H);rs_{--}}_{r_{--},r^2_{--},sr_{--}}=e^{i\frac{2\pi}{3}},\\
        &z^{(H,H);sr_{--}}_{r^2_{--},r_{--},s_{--}}=1,\ z^{(H,H);s_{--}}_{r^2_{--},r_{--},rs_{--}}=e^{i\frac{2\pi}{3}},\ z^{(H,H);rs_{--}}_{r^2_{--},r_{--},sr_{--}}=e^{-i\frac{2\pi}{3}},\\
        &z^{(H,H);\alpha_{+-}}_{r_{++},r^2_{--},\alpha_{+-}}=z^{(H,H);\alpha_{+-}}_{r^2_{++},r_{--},\alpha_{+-}}=z^{(H,H);\beta_{+-}}_{r_{++},r_{--},\beta_{+-}}=z^{(H,H);\beta_{+-}}_{r_{++},r_{--},\beta_{+-}}=e^{-i\frac{2\pi}{3}},\\
        &z^{(H,H);\alpha_{-+}}_{r_{--},r^2_{++},\alpha_{-+}}=z^{(H,H);\alpha_{-+}}_{r^2_{--},r_{++},\alpha_{-+}}=z^{(H,H);\beta_{-+}}_{r_{--},r_{++},\beta_{-+}}=z^{(H,H);\beta_{-+}}_{r^2_{--},r^2_{++},\beta_{-+}}=e^{-i\frac{2\pi}{3}};
    \end{align*}

\begin{align*}
        &z^{(D,D);s_{++}}_{s_{++},s_{++},1_{++}}=z^{(D,D);sr_{++}}_{s_{++},rs_{++},r_{++}}=z^{(D,D);rs_{++}}_{s_{++},sr_{++},r^2_{++}}=1,\\
        &z^{(D,D);rs_{++}}_{rs_{++},rs_{++},1_{++}}=z^{(D,D);s_{++}}_{rs_{++},sr_{++},r_{++}}=z^{(D,D);sr_{++}}_{rs_{++},s_{++},r^2_{++}}=1,\\
        &z^{(D,D);sr_{++}}_{sr_{++},sr_{++},1_{++}}=z^{(D,D);rs_{++}}_{sr_{++},s_{++},r_{++}}=z^{(D,D);s_{++}}_{sr_{++},rs_{++},r^2_{++}}=1,\\
        &z^{(D,D);1_{++}}_{s_{++},s_{++},s_{++}}=z^{(D,D);r_{++}}_{s_{++},rs_{++},sr_{++}}=z^{(D,D);r^2_{++}}_{s_{++},sr_{++},rs_{++}}=1,\\
        &z^{(D,D);1_{++}}_{rs_{++},rs_{++},rs_{++}}=z^{(D,D);r_{++}}_{rs_{++},sr_{++},s_{++}}=z^{(D,D);r^2_{++}}_{rs_{++},s_{++},sr_{++}}=1,\\
        &z^{(D,D);1_{++}}_{sr_{++},sr_{++},sr_{++}}=z^{(D,D);r}_{sr_{++},s_{++},sr_{++}}=z^{(D,D);r^2_{++}}_{sr_{++},rs_{++},s_{++}}=1,\\
        &z^{(D,D);s_{--}}_{s_{--},s_{--},1_{--}}=z^{(D,D);sr_{--}}_{s_{--},rs_{--},r_{--}}=z^{(D,D);rs_{--}}_{s_{--},sr_{--},r^2_{--}}=1,\\
        &z^{(D,D);rs_{--}}_{rs_{--},rs_{--},1_{--}}=z^{(D,D);s_{--}}_{rs_{--},sr_{--},r_{--}}=z^{(D,D);sr_{--}}_{rs_{--},s_{--},r^2_{--}}=1,\\
        &z^{(D,D);sr_{--}}_{sr_{--},sr_{--},1_{--}}=z^{(D,D);rs_{--}}_{sr_{--},s_{--},r_{--}}=z^{(D,D);s_{--}}_{sr_{--},rs_{--},r^2_{--}}=1,\\
        &z^{(D,D);1_{--}}_{s_{--},s_{--},s_{--}}=z^{(D,D);r_{--}}_{s_{--},rs_{--},sr_{--}}=z^{(D,D);r^2_{--}}_{s_{--},sr_{--},rs_{--}}=1,\\
        &z^{(D,D);1_{--}}_{rs_{--},rs_{--},rs_{--}}=z^{(D,D);r_{--}}_{rs_{--},sr_{--},s_{--}}=z^{(D,D);r^2_{--}}_{rs_{--},s_{--},sr_{--}}=1,\\
        &z^{(D,D);1_{--}}_{sr_{--},sr_{--},sr_{--}}=z^{(D,D);R}_{sr_{--},s_{--},sr_{--}}=z^{(D,D);r^2_{--}}_{sr_{--},rs_{--},s_{--}}=1,\\
        &z^{(D,D);\beta_{+-}}_{s_{++},s_{--},\alpha_{+-}}=z^{(D,D);\beta_{+-}}_{s_{++},rs_{--},\alpha_{+-}}=z^{(D,D);\beta_{+-}}_{rs_{++},s_{--},\alpha_{+-}}=z^{(D,D);\beta_{+-}}_{s_{++},sr_{--},\alpha_{+-}}=z^{(D,D);\beta_{+-}}_{sr_{++},s_{--},\alpha_{+-}}=\frac{1}{\sqrt{3}},\\
        &z^{(D,D);\beta_{+-}}_{rs_{++},rs_{--},\alpha_{+-}}=z^{(D,D);\beta_{+-}}_{sr_{++},sr_{--},\alpha_{+-}}=\frac{e^{i\frac{2\pi}{3}}}{\sqrt{3}},\ z^{(D,D);\beta_{+-}}_{rs_{++},sr_{--},\alpha_{+-}}=z^{(D,D);\beta_{+-}}_{sr_{++},rs_{--},\alpha_{+-}}=\frac{e^{-i\frac{2\pi}{3}}}{\sqrt{3}},\\
        &z^{(D,D);\alpha_{+-}}_{s_{++},s_{--},\beta_{+-}}=z^{(D,D);\alpha_{+-}}_{s_{++},rs_{--},\beta_{+-}}=z^{(D,D);\alpha_{+-}}_{rs_{++},s_{--},\beta_{+-}}=z^{(D,D);\alpha_{+-}}_{s_{++},sr_{--},\beta_{+-}}=z^{(D,D);\alpha_{+-}}_{sr_{++},s_{--},\beta_{+-}}=\frac{1}{\sqrt{3}},\\
        &z^{(D,D);\alpha_{+-}}_{rs_{++},rs_{--},\beta_{+-}}=z^{(D,D);\alpha_{+-}}_{sr_{++},sr_{--},\beta_{+-}}=\frac{e^{-i\frac{2\pi}{3}}}{\sqrt{3}},\ z^{(D,D);\alpha_{+-}}_{rs_{++},sr_{--},\beta_{+-}}=z^{(D,D);\alpha_{+-}}_{sr_{++},rs_{--},\beta_{+-}}=\frac{e^{i\frac{2\pi}{3}}}{\sqrt{3}},\\
        &z^{(D,D);\beta_{-+}}_{s_{--},s_{++},\alpha_{-+}}=z^{(D,D);\beta_{-+}}_{s_{--},rs_{++},\alpha_{-+}}=z^{(D,D);\beta_{-+}}_{rs_{--},s_{++},\alpha_{-+}}=z^{(D,D);\beta_{-+}}_{s_{--},sr_{++},\alpha_{-+}}=z^{(D,D);\beta_{-+}}_{sr_{--},s_{++},\alpha_{-+}}=\frac{1}{\sqrt{3}},\\
        &z^{(D,D);\beta_{-+}}_{rs_{--},rs_{++},\alpha_{-+}}=z^{(D,D);\beta_{-+}}_{sr_{--},sr_{++},\alpha_{-+}}=\frac{e^{i\frac{2\pi}{3}}}{\sqrt{3}},\ z^{(D,D);\beta_{-+}}_{rs_{--},sr_{++},\alpha_{-+}}=z^{(D,D);\beta_{-+}}_{sr_{--},rs_{++},\alpha_{-+}}=\frac{e^{-i\frac{2\pi}{3}}}{\sqrt{3}},\\
        &z^{(D,D);\alpha_{-+}}_{s_{--},s_{++},\beta_{-+}}=z^{(D,D);\alpha_{-+}}_{s_{--},rs_{++},\beta_{-+}}=z^{(D,D);\alpha_{-+}}_{rs_{--},s_{++},\beta_{-+}}=z^{(D,D);\alpha_{-+}}_{s_{--},sr_{++},\beta_{-+}}=z^{(D,D);\alpha_{-+}}_{sr_{--},s_{++},\beta_{-+}}=\frac{1}{\sqrt{3}},\\
        &z^{(D,D);\alpha_{-+}}_{rs_{--},rs_{++},\beta_{-+}}=z^{(D,D);\alpha_{-+}}_{sr_{--},sr_{++},\beta_{-+}}=\frac{e^{-i\frac{2\pi}{3}}}{\sqrt{3}},\ z^{(D,D);\alpha_{-+}}_{rs_{--},sr_{++},\beta_{-+}}=z^{(D,D);\alpha_{-+}}_{sr_{--},rs_{++},\beta_{-+}}=\frac{e^{i\frac{2\pi}{3}}}{\sqrt{3}};
\end{align*}

\begin{align*}
        &z^{(E,E);s_{++}}_{s_{++},s_{++},1_{++}}=z^{(E,E);sr_{++}}_{s_{++},rs_{++},r_{++}}=z^{(E,E);rs_{++}}_{s_{++},sr_{++},r^2_{++}}=1,\\
        &z^{(E,E);rs_{++}}_{rs_{++},rs_{++},1_{++}}=z^{(E,E);s_{++}}_{rs_{++},sr_{++},r_{++}}=z^{(E,E);sr_{++}}_{rs_{++},s_{++},r^2_{++}}=1,\\
        &z^{(E,E);sr_{++}}_{sr_{++},sr_{++},1_{++}}=z^{(E,E);rs_{++}}_{sr_{++},s_{++},r_{++}}=z^{(E,E);s_{++}}_{sr_{++},rs_{++},r^2_{++}}=1,\\
        &z^{(E,E);1_{++}}_{s_{++},s_{++},s_{++}}=z^{(E,E);r_{++}}_{s_{++},rs_{++},sr_{++}}=z^{(E,E);r^2_{++}}_{s_{++},sr_{++},rs_{++}}=-1,\\
        &z^{(E,E);1_{++}}_{rs_{++},rs_{++},rs_{++}}=z^{(E,E);r_{++}}_{rs_{++},sr_{++},s_{++}}=z^{(E,E);r^2_{++}}_{rs_{++},s_{++},sr_{++}}=-1,\\
        &z^{(E,E);1_{++}}_{sr_{++},sr_{++},sr_{++}}=z^{(E,E);r}_{sr_{++},s_{++},sr_{++}}=z^{(E,E);r^2_{++}}_{sr_{++},rs_{++},s_{++}}=-1,\\
        &z^{(E,E);s_{--}}_{s_{--},s_{--},1_{--}}=z^{(E,E);sr_{--}}_{s_{--},rs_{--},r_{--}}=z^{(E,E);rs_{--}}_{s_{--},sr_{--},r^2_{--}}=1,\\
        &z^{(E,E);rs_{--}}_{rs_{--},rs_{--},1_{--}}=z^{(E,E);s_{--}}_{rs_{--},sr_{--},r_{--}}=z^{(E,E);sr_{--}}_{rs_{--},s_{--},r^2_{--}}=1,\\
        &z^{(E,E);sr_{--}}_{sr_{--},sr_{--},1_{--}}=z^{(E,E);rs_{--}}_{sr_{--},s_{--},r_{--}}=z^{(E,E);s_{--}}_{sr_{--},rs_{--},r^2_{--}}=1,\\
        &z^{(E,E);1_{--}}_{s_{--},s_{--},s_{--}}=z^{(E,E);r_{--}}_{s_{--},rs_{--},sr_{--}}=z^{(E,E);r^2_{--}}_{s_{--},sr_{--},rs_{--}}=-1,\\
        &z^{(E,E);1_{--}}_{rs_{--},rs_{--},rs_{--}}=z^{(E,E);r_{--}}_{rs_{--},sr_{--},s_{--}}=z^{(E,E);r^2_{--}}_{rs_{--},s_{--},sr_{--}}=-1,\\
        &z^{(E,E);1_{--}}_{sr_{--},sr_{--},sr_{--}}=z^{(E,E);R}_{sr_{--},s_{--},sr_{--}}=z^{(E,E);r^2_{--}}_{sr_{--},rs_{--},s_{--}}=-1,\\
        &z^{(E,E);\beta_{+-}}_{s_{++},s_{--},\alpha_{+-}}=z^{(E,E);\beta_{+-}}_{s_{++},rs_{--},\alpha_{+-}}=z^{(E,E);\beta_{+-}}_{rs_{++},s_{--},\alpha_{+-}}=z^{(E,E);\beta_{+-}}_{s_{++},sr_{--},\alpha_{+-}}=z^{(E,E);\beta_{+-}}_{sr_{++},s_{--},\alpha_{+-}}=\frac{1}{\sqrt{3}},\\
        &z^{(E,E);\beta_{+-}}_{rs_{++},rs_{--},\alpha_{+-}}=z^{(E,E);\beta_{+-}}_{sr_{++},sr_{--},\alpha_{+-}}=\frac{e^{i\frac{2\pi}{3}}}{\sqrt{3}},\ z^{(E,E);\beta_{+-}}_{rs_{++},sr_{--},\alpha_{+-}}=z^{(E,E);\beta_{+-}}_{sr_{++},rs_{--},\alpha_{+-}}=\frac{e^{-i\frac{2\pi}{3}}}{\sqrt{3}},\\
        &z^{(E,E);\alpha_{+-}}_{s_{++},s_{--},\beta_{+-}}=z^{(E,E);\alpha_{+-}}_{s_{++},rs_{--},\beta_{+-}}=z^{(E,E);\alpha_{+-}}_{rs_{++},s_{--},\beta_{+-}}=z^{(E,E);\alpha_{+-}}_{s_{++},sr_{--},\beta_{+-}}=z^{(E,E);\alpha_{+-}}_{sr_{++},s_{--},\beta_{+-}}=\frac{1}{\sqrt{3}},\\
        &z^{(E,E);\alpha_{+-}}_{rs_{++},rs_{--},\beta_{+-}}=z^{(E,E);\alpha_{+-}}_{sr_{++},sr_{--},\beta_{+-}}=\frac{e^{-i\frac{2\pi}{3}}}{\sqrt{3}},\ z^{(E,E);\alpha_{+-}}_{rs_{++},sr_{--},\beta_{+-}}=z^{(E,E);\alpha_{+-}}_{sr_{++},rs_{--},\beta_{+-}}=\frac{e^{i\frac{2\pi}{3}}}{\sqrt{3}},\\
        &z^{(E,E);\beta_{-+}}_{s_{--},s_{++},\alpha_{-+}}=z^{(E,E);\beta_{-+}}_{s_{--},rs_{++},\alpha_{-+}}=z^{(E,E);\beta_{-+}}_{rs_{--},s_{++},\alpha_{-+}}=z^{(E,E);\beta_{-+}}_{s_{--},sr_{++},\alpha_{-+}}=z^{(E,E);\beta_{-+}}_{sr_{--},s_{++},\alpha_{-+}}=\frac{1}{\sqrt{3}},\\
        &z^{(E,E);\beta_{-+}}_{rs_{--},rs_{++},\alpha_{-+}}=z^{(E,E);\beta_{-+}}_{sr_{--},sr_{++},\alpha_{-+}}=\frac{e^{i\frac{2\pi}{3}}}{\sqrt{3}},\ z^{(E,E);\beta_{-+}}_{rs_{--},sr_{++},\alpha_{-+}}=z^{(E,E);\beta_{-+}}_{sr_{--},rs_{++},\alpha_{-+}}=\frac{e^{-i\frac{2\pi}{3}}}{\sqrt{3}},\\
        &z^{(E,E);\alpha_{-+}}_{s_{--},s_{++},\beta_{-+}}=z^{(E,E);\alpha_{-+}}_{s_{--},rs_{++},\beta_{-+}}=z^{(E,E);\alpha_{-+}}_{rs_{--},s_{++},\beta_{-+}}=z^{(E,E);\alpha_{-+}}_{s_{--},sr_{++},\beta_{-+}}=z^{(E,E);\alpha_{-+}}_{sr_{--},s_{++},\beta_{-+}}=\frac{1}{\sqrt{3}},\\
        &z^{(E,E);\alpha_{-+}}_{rs_{--},rs_{++},\beta_{-+}}=z^{(E,E);\alpha_{-+}}_{sr_{--},sr_{++},\beta_{-+}}=\frac{e^{-i\frac{2\pi}{3}}}{\sqrt{3}},\ z^{(E,E);\alpha_{-+}}_{rs_{--},sr_{++},\beta_{-+}}=z^{(E,E);\alpha_{-+}}_{sr_{--},rs_{++},\beta_{-+}}=\frac{e^{i\frac{2\pi}{3}}}{\sqrt{3}}.
\end{align*}

\section{Anyon Fragmentation Pattern}\label{appendix:frag}

In this section, we provide the non-Abelian anyon fragmentation pattern of $D(S_3)$ phase under EM-exchange symmetry (in $\alpha$-gauge), as shown in Table \ref{tab:frag}.

\begin{table}[h]
    \centering
    \begin{tabular}{|c|c|c|}
        \hline
        \textbf{Non-Abelian anyons} & \textbf{Definite symmetry charge states (spaces)} & \textbf{Symmetry charge} \\
        \hline
        \multirow{2}{*}{$C\oplus F$} & ${\tt span}\{C_{1_1}+ F_r,\ C_{1_2}+ F_{r^2}\}$ & $0$ \\
        \cline{2-3}
        & ${\tt span}\{C_{1_1}- F_r,\ C_{1_2}- F_{r^2}\}$ & $\frac{1}{2}$ \\
        \hline
        $G$ & ${\tt span}\{G_{r},\ G_{r^2}\}$ & $\frac{1}{3}$ \\
        \hline
        \multirow{2}{*}{$H$} & $H_r+H_{r^2}$ & $\frac{2}{3}$ \\
        \cline{2-3}
        & $H_r-H_{r^2}$ & $\frac{1}{6}$ \\
        \hline
        \multirow{3}{*}{$D$} & $D_{rs}-D_{sr}$ & $\frac{1}{4}$ \\
        \cline{2-3}
        & $(1+\sqrt{3})D_s+D_{rs}+D_{sr}$ & $0$ \\
        \cline{2-3}
        & $(1-\sqrt{3})D_s+D_{rs}+D_{sr}$ & $\frac{1}{2}$ \\
        \hline
        \multirow{3}{*}{$E$} & $E_{rs}-E_{sr}$ & $\frac{1}{4}$ \\
        \cline{2-3}
        & $(1+\sqrt{3})E_s+E_{rs}+E_{sr}$ & $0$ \\
        \cline{2-3}
        & $(1-\sqrt{3})E_s+E_{rs}+E_{sr}$ & $\frac{1}{2}$ \\
        \hline
    \end{tabular}
    \caption{Non-Abelian anyon fragmentation pattern of $D(S_3)$ phase under EM-exchange symmetry (in $\alpha$-gauge)}
    \label{tab:frag}
\end{table}

\section{Proof that $\rho^G$ and $\rho^H$ are not projective representations}\label{app:proof}

Here, we prove that the representations $\rho^G$ \eqref{eq:Hrep} and $\rho^H$ \eqref{eq:Grep} are not projective representations, which are equivalent to linear representations of $\Z_2$, although they appear to be. Our proof goes by showing that the phase factors in \eqref{eq:Hrep} and \eqref{eq:Grep}, which are due to the nonlinearity indicator $\omega$, cannot be transformed away consistently:

Since the nonlinearity indicator $\omega$ only depends on the Pachner moves, we can express the components of $\omega$ solely using $6j$-symbols as
\begin{equation}\label{eq:omega}
    \begin{split}
        & \omega_{r,r}=3G^{r^2,\alpha,\alpha}_{r,\alpha,\alpha}G^{1,r,r^2}_{\alpha,\alpha,\alpha}G^{1,r,r^2}_{\alpha,\alpha,\alpha},\\
        & \omega_{r^2,r^2}=3G^{r,\alpha,\alpha}_{r^2,\alpha,\alpha}G^{1,r,r^2}_{\alpha,\alpha,\alpha}G^{1,r,r^2}_{\alpha,\alpha,\alpha},\\
        & \omega_{r,r^2}=3G^{r^2,\alpha,\alpha}_{r^2,\alpha,\alpha}G^{1,r,r^2}_{\alpha,\alpha,\alpha}G^{1,r,r^2}_{\alpha,\alpha,\alpha},\\
        & \omega_{r^2,r}=3G^{r,\alpha,\alpha}_{r,\alpha,\alpha}G^{1,r,r^2}_{\alpha,\alpha,\alpha}G^{1,r,r^2}_{\alpha,\alpha,\alpha}.\\
    \end{split}
\end{equation}
(Here, as $G^{a_{ij}b_{jk}e_{ik}}_{c_{kl}d_{li}f_{jl}}=G^{a_{\bar{i}\bar{j}}b_{\bar{j}\bar{k}}e_{\bar{i}\bar{k}}}_{c_{\bar{k}\bar{l}}d_{\bar{l}\bar{i}}f_{\bar{j}\bar{l}}}$ always holds for any $i,j,k,l\in\{+,-\}$ due to the symmetry, we use $G^{abe}_{cdf}$ to denote them for convenience: $G^{abe}_{cdf}:=G^{a_{ij}b_{jk}e_{ik}}_{c_{kl}d_{li}f_{jl}}=G^{a_{\bar{i}\bar{j}}b_{\bar{j}\bar{k}}e_{\bar{i}\bar{k}}}_{c_{\bar{k}\bar{l}}d_{\bar{l}\bar{i}}f_{\bar{j}\bar{l}}}$.) In order to transform these $\omega$ away, we can only resort to the gauge transformation on the $6j$-symbols\cite{hung2012,fu2025symmetryenrichedtopologicalphasesgauging}
\begin{equation}\label{eq:gaugetrans}
    G^{abe}_{cdf}\to \frac{u_{abe}u_{cde^*}}{u_{daf^*}u_{bcf}}G^{abe}_{cdf}.
\end{equation}
But this is impossible without violating the gauge constraint condition
\begin{equation}
    G^{0,\alpha,\alpha}_{i,\alpha,\alpha}=1,\ \forall i\in\{1,r,r^2\},
\end{equation}
which must be satisfied by the string-net model with any input UFC.

Thus, the phase factors in \eqref{eq:Hrep} \eqref{eq:Grep} cannot be transformed away consistently in the string-net model, rendering $\rho^G$ and $\rho^H$ truly nonlinear representations.

\bibliographystyle{apsrev4-1}
\bibliography{StringNet}
\end{document}